\documentclass[twocolumn]{revtex4-2}
\usepackage[dvipdfmx]{graphicx}
\usepackage{amsmath,amsbsy,amssymb}
\usepackage{bm}
\usepackage{mathrsfs}
\usepackage{ulem}
\usepackage{textgreek}
\usepackage{mathrsfs}
\usepackage{mleftright}
\usepackage{braket}
\usepackage{physics}
\usepackage{hyperref}

\usepackage{color}

\newcommand{\imu}{\mathrm{i}}
\newcommand{\epn}{\mathrm{e}}

\newcommand{\ua}{\uparrow}
\newcommand{\da}{\downarrow}
\newcommand{\dg}{\dagger}
\newcommand{\la}{\langle}
\newcommand{\ra}{\rangle}
\newcommand{\al}{\alpha}
\newcommand{\sg}{\sigma}
\newcommand{\gm}{\gamma}
\newcommand{\ep}{\varepsilon}

\begin{document}

\title{
Spin-orbital model for fullerides
}

\author{
Ryuta Iwazaki and Shintaro Hoshino
}

\affiliation{
Department of Physics, Saitama University, Shimo-Okubo, Saitama 338-8570, Japan
}

\date{\today}

\begin{abstract}
The multiorbital Hubbard model in the strong coupling limit is analyzed for the effectively antiferromagnetic Hund's coupling relevant to fulleride superconductors with three orbitals per molecule.
The localized spin-orbital model describes the thermodynamics of the half-filled (three-electron) state with total spin-1/2, composed of singlon and doublon placed on the two of three orbitals.
The model is solved using the mean-field approximation and magnetic and electric ordered states are clarified through the temperature dependences of the order parameters.
Combining the model with the band structure from {\it ab initio} calculation, we also semi-quantitavely analyze 
the realistic model and the corresponding physical quantities.
In the A15-structure fulleride model, there is an antiferromagnetic ordered state, and subsequently the two orbital ordered state appears at lower temperatures.	It is argued that the origin of these orbital orders is related to the $T_h$ point group symmetry.
As for the fcc-fulleride model, the time-reversal broken orbital ordered state is identified. Whereas the spin degeneracy remains in our treatment for the geometrically frustrated lattice, it is expected to be lifted by some magnetic ordering or quantum fluctuations, but not by the spin-orbital coupling which is effectively zero for fullerides in the strong-coupling regime.
\end{abstract}

\maketitle

\section{Introduction}
	Strongly correlated electron systems with multiple orbital degrees of freedom show a variety of intriguing phenomena, and are realized in a wide range of materials such as iron-pnictides, heavy-electron materials, and molecular-based organic materials.
The alkali-doped fullerides are also the typical cases where the strong correlation effects with multiorbitals are relevant.
This material has been attracting attention recent years for a lot of experimental  findings.
The superconductivity with the high transition temperature $\sim 40$K is one of the characteristic feature~\cite{Hebard1991, Rosseinsky1991, Holczer1991, Tanigaki1991, Fleming1991, Ganin2008, Takabayashi2009, Ganin2010}.
While the mechanism is identified as the electron-phonon interaction~\cite{Gunnarsson1997, Fabrizio1997, Capone2002}, the superconducting dome in the temperature-pressure phase diagram is found to be located near the Mott insulator and antiferromagnetic phase, featuring the typical behaviors of the strongly correlated superconductors~\cite{Takabayashi2009, Capone2009, Nomura2016, Takabayashi2016}.
In the Mott insulating phase, the localized electrons form a low-spin state and the imbalance of the occupancy in orbitals lead to the deformation of the fullerene molecule because of the coupling between electrons and anisotropic molecular distortions (Jahn-Teller phonon).
Interestingly, such behavior can also be seen in the metallic phase near the Mott insulator but is absent far away from it~\cite{Zadik2015, Kasahara2017}.
This anomalous behavior is called the Jahn-Teller metal where the multiorbital degrees of freedom play an important role.
The fullerides are also crystallized on the substrate and the characteristic asymmetry between electron and hole doping is identified~\cite{Han2020, Ren2020}.
Furthermore, a possible superconducting state has been discussed under the excitation by light above the transition temperature~\cite{Mitrano2016, Cantaluppi2018}.
Thus, the fulleride materials have been providing the intriguing phenomena up until recently.

The alkali-doped fullerides are the systems with triply degenerate $t_{1u}$ molecular orbitals which resembles atomic $p$-electrons in nature.
There, the Hund's coupling, which is usually acting ferromagnetically on the electrons located at the different orbitals, is effectively antiferromagnetic due to the coupling to the anisotropic molecular vibrations~\cite{Fabrizio1997, Capone2000, Nomura2015sci} and is crucial for the low-temperature physics.
The multiorbital Hubbard model with the antiferromagnetic Hund's coupling has been studied theoretically, and the various phase diagrams are clarified using the dynamical mean-field theory suitable for the description of the electronically ordered states~\cite{Capone2000, Koga2015, Nomura2015sci, Hoshino2016, Steiner2016, Hoshino2017, Ishigaki2018, Ishigaki2019, Yue2020nov}.
The Jahn-Teller metal has been interpreted as the spontaneous orbital selective Mott state~\cite{Hoshino2017, Hoshino2019} which is an unconventional type of orbital order.
The orbital asymmetric feature has also been reported in two-dimensional fullerides by using the many-variable variational Monte Carlo method~\cite{Misawa2017}.

With the antiferromagnetic Hund's coupling, one of the intra-molecular interaction, pair hopping, plays an important role: it activates the dynamics of the double occupancy in an orbital (doublon).
In order to clarify the characters of the existing fulleride materials in detail, we focus our attention on the Mott insulating phase, where the doublon physics can be tackled with reasonable computational cost even in the realistic situation. 
As is well known, for a single-orbital case, the electronic behaviors in the strong coupling regime are determined by the Heisenberg model of localized electrons.
The extension of the Heisenberg model to the multiorbital system is known as the Kugel-Khomskii model which has been derived for the ferromagnetic Hund's coupling~\cite{Kugel72, Kugel73} and describes the degrees of freedom of the spin and orbital.
The spin-orbital models have been applied to the $e_g$ or $t_{2g}$ orbital system~\cite{Inagaki1975, Ishihara1997, Feiner1999, Normand2008}.
On the other hand, the fullerides have antiferromagnetic Hund's coupling, so that their strongly correlated effective model differs from the usual Kugel-Khomskii model.
While the localized model with antiferromagnetic Hund's coupling have been constructed for a density-density type interaction~\cite{Ishigaki2019}, here we deal with more complicated but realistic situations.

In this paper, we develop the localized spin-orbital model for the system with antiferromagnetic Hund's coupling.
We analyze both the symmetric model and the realistic model for fullerides, the former of which is easier to interpret the results and is useful as a reference.
By using the mean field theory, for the spherical model on a bipartite lattice, we obtain the staggered magnetic ordered state, and also the uniform orbital ordered state at lower temperature regime.
This orbital ordered state is not characterised by the ordinary orbital moment but by the doublon's orbital moment.
In the A15 fulleride effective model, which is bipartite lattice, we reveal that there are two kinds of orbital ordered states below the antiferromagnetic transition temperature.
The obtained orbital ordered states are interpreted as related to an effective recovery of the four-fold symmetry at low temperatures in the $T_h$ point group.
We also analyze the geometrically frustrated fcc fulleride model seeking for a spatially uniform ordered state.
We reveal that the fcc model has the time-reversal symmetry broken orbital ordered state, where the spin ordered state is absent since the spin-orbit coupling on the fullerene molecule is effectively zero.

	This paper is organized as follows.
	We discuss the construction of strongly correlated effective models and the theoretical method in Sec.~\ref{sec:models}.
	In Sec.~\ref{sec:results_spherical}, we show numerical results for the model with isotropic hopping (spherical model introduced in Sec.~\ref{sec:effective_model}).
	Section~\ref{sec:results_fullerides} provides numerical results for the spin-orbital model combined with A15 and fcc fulleride band structure.
	We summarize the results in Sec.~\ref{sec:summary}.

\section{Construction of models} \label{sec:models}
\subsection{Three orbital Hubbard model in strong-coupling limit} \label{sec:hubbard}
    Let us begin with the three-orbital Hubbard model
    \begin{align}
        &\mathcal{H}
        =
        \mathcal{H}_t + \mathcal{H}_U,
        \label{eq:total_hamiltonian}
        \\
        &\mathcal{H}_t
        =
		-\sum_{i \neq j, \gm, \gm', \sg} t_{ij}^{\gm\gm'} 
		c_{i, \gm, \sg}^{\dg} c_{j, \gm', \sg},
        \label{eq:H_t}
		\\
		&\mathcal{H}_U
		=
		\frac{U}{2} \sum_{i, \gm, \sg, \sg'} c_{i, \gm,\sg}^{\dg} c_{i, \gm,\sg'}^{\dg} c_{i, \gm,\sg'} c_{i, \gm,\sg}
		\nonumber
		\\
		&\hspace{.7cm}
		+
		\frac{U'}{2} \sum_{i, \gm\neq\gm', \sg, \sg'} c_{i, \gm,\sg}^{\dg} c_{i, \gm',\sg'}^{\dg} c_{i, \gm',\sg'} c_{i, \gm,\sg}
		\nonumber
		\\
		&\hspace{.7cm}
		+
		\frac{J}{2} \sum_{i, \gm\neq\gm', \sg, \sg'} \Big(
			c_{i, \gm,\sg}^{\dg} c_{i, \gm',\sg'}^{\dg} c_{i, \gm,\sg'} c_{i, \gm',\sg}
			\nonumber
			\\
			&\hspace{3.7cm}
			+
			c_{i, \gm,\sg}^{\dg} c_{i, \gm,\sg'}^{\dg} c_{i, \gm',\sg'} c_{i, \gm,\sg}
		\Big),
        \label{eq:H_U}
    \end{align}
    where $c_{i, \gm, \sg}$ ($c_{i, \gm, \sg}^\dg$) is an annihilation (creation) operator at site $i$ of fullerenes with the $t_{1u}$ molecular orbital index $\gm=x,y,z$ and spin $\sg=\ua,\da$.
    We deal with the Hilbert space with a fixed number of electrons.
    We assume the condition $U' = U - 2J$ for the local interaction part in the following discussion, which is valid for the spherical limit.
    In this paper, we consider a strong coupling regime ($\mathcal{H}_U \gg \mathcal{H}_t$).
    When we develop the effective model in this limit, the presence of the Hund's coupling  $J$ makes theoretical treatment complicated since it realizes quantum-mechanically superposed local wave functions.
    Especially for the negative (antiferromagnetic) $J$ relevant to fullerides, the pair hopping plays an important role which creates the dynamics of doubly occupied electrons at an orbital (doublon).
    As shown in the following, in order to diminish the difficulty, we use a symbolic expression without elaborating each intermediate process explicitly.

    \begin{figure}[t]
        \centering
        \includegraphics[width=85mm]{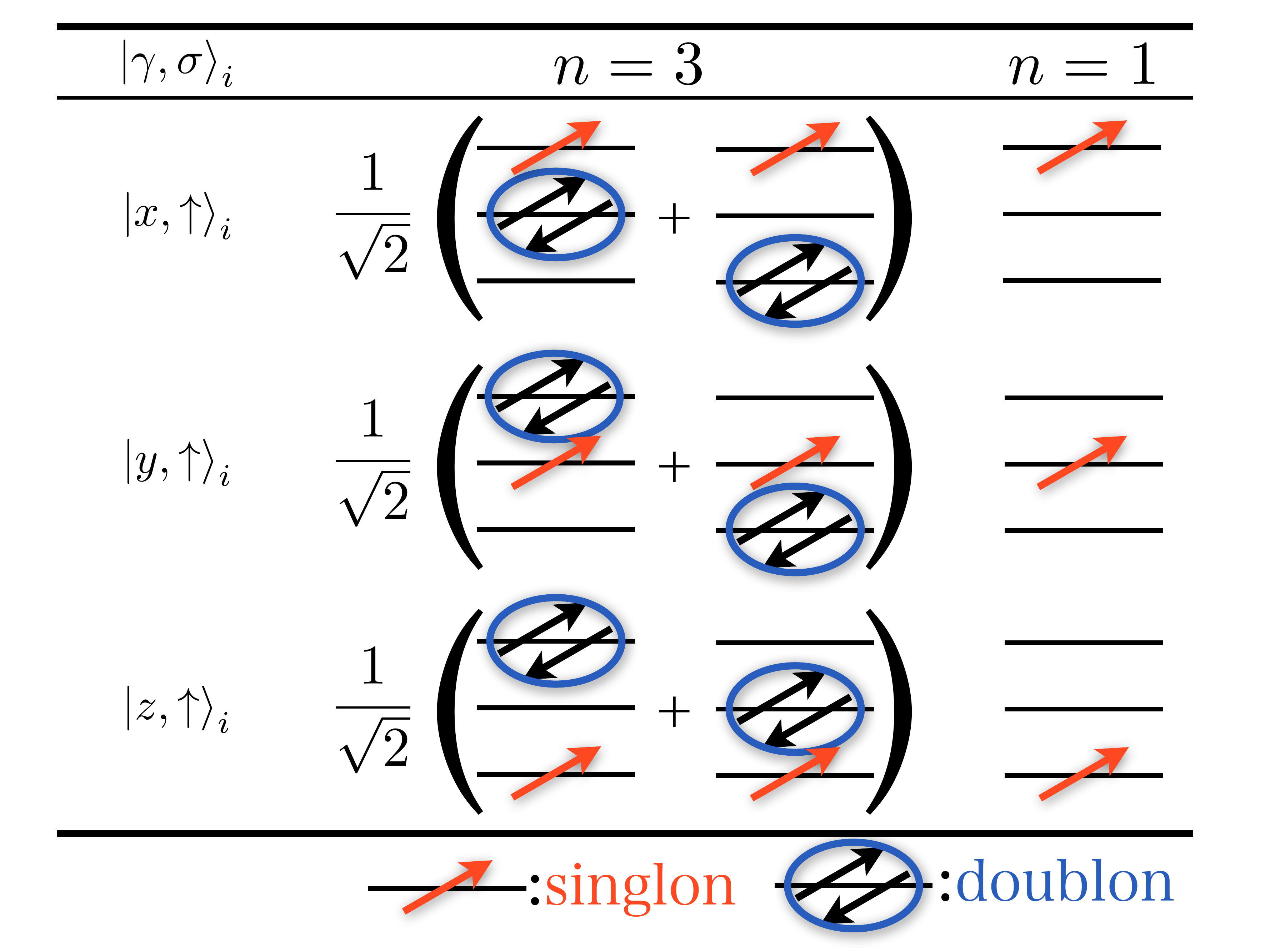}
        \caption{
            Schematic pictures for the ground state wave functions $\Ket{\gm, \sg = \ua}_i$ of the local Hamiltonian for $n = 3$ and $n = 1$.
        }
        \label{fig:ground_state}
    \end{figure}
    
    In order to apply the perturbation theory from the strong coupling limit, we first consider the ground state of the unperturbed Hamiltonian $\mathcal{H}_U$.
    Alkali-doped fullerides with half-filled situation (three electrons per $t_{1u}$ orbital) have six-fold degenerate ground states written as
    \begin{align} \label{eq:ground_state_single_site}
        \ket{\gm, \sg}_i
        =
        \frac{1}{\sqrt{2}} c_{i, \gm, \sg}^{\dg} \sum_{\gm' \neq \gm} b_{i, \gm'}^{\dg} \ket{0},
    \end{align}
    where we have defined an orbital-dependent doublon-creation operator as
    \begin{align} \label{eq:def_operator_b}
        b_{i, \gm}^{\dg}
        =
        c_{i, \gm, \da}^{\dg} c_{i, \gm, \ua}^{\dg}.
    \end{align}
The vacuum has been expressed as $\Ket{0}$.
These states are uniquely characterized by the spin and orbital of the electron at the singly occupied orbital, which is called `singlon' to make contrast against doublons.
The schematic picture of the three-electron state $\Ket{\gm, \sg=\ua}_i$ is illustrated in Fig.~\ref{fig:ground_state}.

    Using the above Hamiltonian, the second-order effective Hamiltonian is written as
    \begin{align}
        \label{eq:H_eff}
        \mathcal{H}_{\mathrm{eff}}
        =
        \mathcal{P} \mathcal{H}_t \frac{1}{- \mathcal{H}_U} \mathcal{Q} \mathcal{H}_t \mathcal{P},
    \end{align}
    where $\mathcal{P}$ is a projection operator to a model space described by Eq.~\eqref{eq:ground_state_single_site} as
\begin{align}
\mathcal P &= \sum_{i,\gm,\sg} |\gm, \sg \ra_i \,_i \la \gm, \sg |,
\end{align}
and $\mathcal{Q} = 1 -\mathcal P$.
We have used $\comm{\mathcal{P}}{\mathcal{H}_U}$ = 0.
The energy is measured from the ground state of $\mathcal H_U$.
The size of our model space is $6^N$ where $N=\sum_i 1$ is the number of lattice sites.

    The strategy for obtaining the concrete form of the effective Hamiltonian is to consider the two-site problem.
    We first prepare the $2^{12}\times 2^{12}$ matrix expressions for the annihilation and creation operators for two-site problem ($12 = \sum_{i,\gm,\sg}1$), and then define all of the matrix expressions given in Eq.~\eqref{eq:H_eff}.
    Performing multiplications of such matrices, we obtain the two-site effective Hamiltonian in the form of the $6^2 \times 6^2$ matrix.
    We expand the above effective hamiltonian by following local operators $O_i^{\eta \mu}$ defined as
    \begin{align}
        O_i^{\eta \mu}
        &=\sum_{\gm,\gm'} \sum_{\sg,\sg'} |\gm, \sg \ra_i \lambda^{\eta}_{\gm\gm'} \sg^\mu_{\sg\sg'} \,_i \la \gm', \sg' |,
        \label{eq:def_operator_O}
    \end{align}
in the model Hilbert space.
$\sg^{\mu=0, x,y,z}$ is Pauli matrix 
\begin{align}
&\sg^0 = \begin{pmatrix}
1 & 0 \\
0 & 1
\end{pmatrix}
, \ \ 
\sg^x = \begin{pmatrix}
0 & 1 \\
1 & 0
\end{pmatrix}
, \nonumber \\
&\sg^y = \begin{pmatrix}
0 & -\imu \\
\imu & 0
\end{pmatrix}
, \ \ 
\sg^z = \begin{pmatrix}
1 & 0 \\
0 & -1
\end{pmatrix}
, \ \ 
\end{align}
which represents the degrees of freedom of the spin.
Another matrix $\lambda^{\eta=0,\cdots,8}$ is given by
    \begin{align}
        &\lambda^0
        =
        \sqrt{\frac{2}{3}} \mqty(
            1 & 0 & 0 \\
            0 & 1 & 0 \\
            0 & 0 & 1
        ),
        \hspace{.5cm}
        \lambda^1
        =
        \mqty(
            0 & -1 & 0 \\
            -1 & 0 & 0 \\
            0 & 0 & 0
        ),
        \nonumber
        \\
        &\lambda^2
        =
        \mqty(
            0 & -\imu & 0 \\
            \imu & 0 & 0 \\
            0 & 0 & 0
        ),
        \hspace{.5cm}
        \lambda^3
        =
        \mqty(
            -1 & 0 & 0 \\
            0 & 1 & 0 \\
            0 & 0 & 0
        ),
        \nonumber
        \\
        &\lambda^4
        =
        \mqty(
            0 & 0 & -1 \\
            0 & 0 & 0 \\
            -1 & 0 & 0
        ),
        \hspace{.5cm}
        \lambda^5
        =
        \mqty(
            0 & 0 & \imu \\
            0 & 0 & 0 \\
            -\imu & 0 & 0
        ),
        \nonumber
        \\
        &\lambda^6
        =
        \mqty(
            0 & 0 & 0 \\
            0 & 0 & -1 \\
            0 & -1 & 0
        ),
        \hspace{.5cm}
        \lambda^7
        =
        \mqty(
            0 & 0 & 0 \\
            0 & 0 & -\imu \\
            0 & \imu & 0
        ),
        \nonumber
        \\
        &\lambda^8
        =
        \sqrt{\frac{1}{3}} \mqty(
            1 & 0 & 0 \\
            0 & 1 & 0 \\
            0 & 0 & -2
        ),
    \end{align}
    where these matrices are slightly different from ordinary definition of the Gell-Mann matrices to make them suitable for $p$-electron systems.
    We note that the above local operators satisfy the orthonormal relation
    \begin{align}
        \Tr \qty[O_i^{\eta\mu} O_j^{\eta'\mu'}]
        =
        4\delta_{ij} \delta^{\eta\eta'} \delta^{\mu\mu'}.
    \end{align}
    Thus, the set of operators $O_i^{\eta\mu}$ is regarded as a basis set of the extended Hilbert space (Liouville space).
In contrast, the states $|\gm, \sg\ra_i$ are the basis in the six-component model Hilbert space.
Extending the two-site problem to the full lattice, we obtain the effective Hamiltonian in the strong coupling limit
\begin{align}
\mathcal H_{\rm eff} &= \sum_{i,j} \sum_{\eta,\eta'} \sum_{\mu,\mu'} I_{ij}^{\eta\mu;\eta'\mu'}
O_i^{\eta\mu} O_j^{\eta'\mu'}.
\end{align}
This model is to be analyzed in the rest of this paper.

We also comment on the orbital moments in the restricted Hilbert space.
In terms of the original Hubbard model, the local orbital moment is defined by
\begin{align}
\bm {\mathcal L}_i &\equiv \sum_{\gm,\gm',\sg} c^\dg_{i,\gm,\sg} \bm \ell_{\gm\gm'} c_{i,\gm',\sg},
\end{align}
where the 3$\times $3 matrices are given by $\ell_x = \lambda^{7}$,  $\ell_y = \lambda^{5}$, and  $\ell_z = \lambda^{2}$.
This angular momentum operator is, however, zero for the restricted Hilbert space: 
\begin{align}
\mathcal P \bm {\mathcal L}_i \mathcal P = \bm 0.
\end{align}
This anomalous disappearance of the angular momentum is due to the composite nature of the ground state~\cite{Hoshino2017} and is very different from a singly occupied state.
Then the active orbital degrees of freedom are not of the original electrons but of the three-electron composite involving doublons.
This feature also affects the spin-orbit coupling which takes the form
\begin{align}
\mathcal H_{\rm SO} &= \frac 1 2 \lambda_{\rm SO} \sum_i
\sum_{\gm,\gm'}\sum_{\sg,\sg'} c^\dg_{i,\gm,\sg} \bm \ell_{\gm\gm'} \cdot \bm \sg_{\sg\sg'} c_{i,\gm',\sg'},
\end{align}
in the language of the original multiorbital Hubbard model.
The spin-orbit coupling for $2p$-electron in carbon atom is nearly $2$meV, and because of the extended nature of the fullerene molecular the spin-orbit coupling $\lambda_{\rm SO}$ for $t_{1u}$ orbitals is one-hundred times smaller than the atomic value ($\lambda_{\rm SO} \sim 20\mu$eV)~\cite{Tosatti1996}.
Furthermore, for the restricted Hilbert space of $n=3$ states, the effect of the spin-orbit coupling enters only through the second-order perturbation contribution as
\begin{align}
\mathcal H_{\rm SO}^{(2)} &= \mathcal P \mathcal H_{\rm SO} \frac{1}{-\mathcal H_U} \mathcal Q \mathcal H_{\rm SO} \mathcal P
\\
&= \frac 1 2 \Lambda_{\rm SO}\sum_i
\sum_{\gm,\gm'}\sum_{\sg,\sg'} |\gm, \sg\ra_i \bm \ell_{\gm\gm'} \cdot \bm \sg_{\sg\sg'} \,_{i}\la \gm', \sg'|,
\end{align}
where $\Lambda_{\rm SO} = \frac{11 \lambda_{\rm SO}^2}{20 J}$ for $J<0$.
Using the values for the antiferromagnetic coupling $J \sim - 0.03$eV for fullerdies~\cite{Nomura2015prb}, we obtain $\Lambda_{\rm SO} \sim 1$neV which is so tiny.
Hence we can safely neglect the spin-orbit coupling in fullerides.

It is convenient to recognize that the above three electron state is similar to the singly occupied state
\begin{align}
\Ket{n=1,\gm, \sg}_i &= c^\dg_{i,\gm,\sg} \Ket{0},
\label{eq:wf_n=1}
\end{align}
which is the eigenstate with $n_i = \sum_{\gm, \sg} c_{i, \gm, \sg}^{\dg} c_{i, \gm, \sg}=1$ regardless of the sign of $J$ (see the right column of Fig.~\ref{fig:ground_state}).
In our paper, the number of electrons is fixed at each site and $n_i$ is sometimes 
simply written as $n$.
In Eq.~\eqref{eq:wf_n=1}, we explicitly write `$n=1$', and if it is dropped, the state represents $n=3$ state defined in Eq.~\eqref{eq:ground_state_single_site}.
    The ground state for $n = 3$ is obtained by filling the empty orbital in $n=1$ state by the doublons as in
Eq.~\eqref{eq:ground_state_single_site}.

We will consider the $n=1$ case for reference to illuminate the characteristics of $n=3$ relevant to fullerides.
When we deal with the second-order effective Hamiltonian for the $n=1$ states, we just replace $\Ket{\gm, \sg}_i$ by $\Ket{n=1,\gm, \sg}_i$ defined in Eq.~\eqref{eq:wf_n=1}.
We note that, in this case, the angular momentum does not vanish as distinct from the $n=3$ multiplet.
For the usual ferromagnetic Hund's coupling ($J>0$), the system corresponds to the spin-orbital model considered for the $t_{2g}$ orbitals~\cite{Normand2008}.

\subsection{Mean field approximations} \label{sec:mfa_procedure}
    In this paper, we utilize the mean field approximation (MFA) for the obtained effective Hamiltonian.
We apply the external field for convenience and the full Hamiltonian is written as
    \begin{align}
        \mathcal{H}_{\mathrm{eff}}
        =&
        \frac{1}{2} \sum_{i,j} \vec{O}_i^{\mathrm{T}} \hat{I}_{ij} \vec{O}_j
        -
        \sum_i \vec{H}_i^{\mathrm{T}} \vec{O}_i
        \\
        \approx&
        -\sum_{i,j} \qty[
            \vec{H}_i^{\mathrm{T}} {\delta}_{ij} \hat 1 
            -
            \frac{1}{2} \vec{\mathcal{M}}_i^{\mathrm{T}} \qty(\hat{I}_{ij} + \hat{I}_{ji}^{\mathrm{T}})
        ] \vec{O}_j
        \nonumber
        \\
        &-
        \frac{1}{2} \sum_{i,j} \vec{\mathcal{M}}_i^{\mathrm{T}} \hat{I}_{ij} \vec{\mathcal{M}}_j
\equiv \mathcal H^{\rm MF},
    \end{align}
    where the hat and arrow symbols represent the matrix and vector, respectively, with respect to the intra-site degrees of freedom $(\eta,\mu)$. 
    The vector $\vec{O}_i$ is the operator for the order parameter at site $i$, whose matrix representation is given in Eq.~\eqref{eq:def_operator_O}.
    Namely, it is a column vector having 35 components, each of which is a 6$\times$6 matrix where the identity is eliminated.
    The statistical average $\vec{\mathcal{M}}_i = \la \vec O_i \ra$ is the order parameter.
    In this paper, the coupling constant $\hat{I}_{ij}$ connects only nearest-neighbor (NN) sites for the spherical model (Sec.~\ref{sec:results_spherical}), and NN and next-nearest-neighbor (NNN) site for A15 and fcc fulleride model (Sec.~\ref{sec:results_fullerides}).
    In the following of this section, we concentrate on the bipartite lattice such as A15 structure.
    Then we introduce two kinds of AB-sublattice to describe staggered orders.
    For non-bipartite lattice (i.e. fcc), on the other hand, we consider only the uniform solution and the similar formula can easily be obtained by regarding the two sublattices as identical.

The mean-field Hamiltonian is then rewritten as
    \begin{align} \label{eq:MF_hamiltonian_sub-lattice}
        &\mathcal{H}^{\mathrm{MF}}
        =
        -\sum_{\al} \Bigg[
            \vec{H}_{\al}^{\mathrm{T}}
            -
            \frac{1}{2} \sum_{\delta\in \mathrm{NN}} \vec{\mathcal{M}}_{\bar{\al}}^{\mathrm{T}} \qty(\hat{I}_{\delta,0} + \hat{I}_{0,\delta}^{\mathrm{T}})
            \nonumber
            \\
            &-
            \frac{1}{2} \sum_{\delta\in \mathrm{NNN}} \vec{\mathcal{M}}_{\al}^{\mathrm{T}} \qty(\hat{I}_{\delta,0} + \hat{I}_{0,\delta}^{\mathrm{T}})
        \Bigg] \sum_{i\in\al}^{N/2} \vec{O}_i
        \nonumber
        \\
        &-
        \frac{1}{2} \frac{N}{2} \sum_{\al} \qty[
            \sum_{\delta\in \mathrm{NN}} \vec{\mathcal{M}}_{\bar{\al}}^{\mathrm{T}} \hat{I}_{\delta, 0} \vec{\mathcal{M}}_{\al}
            +
            \sum_{\delta\in \mathrm{NNN}} \vec{\mathcal{M}}_{\al}^{\mathrm{T}} \hat{I}_{\delta, 0} \vec{\mathcal{M}}_{\al}
        ],
    \end{align}
    where $\al = {\rm A,B}$ is the sub-lattice index and $\bar{\al}$ is a complementary component of $\al$, i.e., ${\rm \bar A} = {\rm B}$ and ${\rm \bar B} = {\rm A}$.
    $N$ is the number of site.
    The number of $\delta\in \mathrm{NN}$ is $z$, $8$ or $12$ respectively for the spherical, A15 or fcc model.
    As for $\delta\in \mathrm{NNN}$, both the A15 case (and fcc) has six sites.
We have used the fact that NN-connected sites belong to the different sub-lattices and the NNN-connected sites belongs to the same sub-lattice.
    Since the coupling constants are dependent only on the direction of the vector connecting two sites, we write the interaction parameter as $\hat{I}_{\delta,0}$, where the index 0 represents the site which we focus on.

        For the bipartite lattice, we introduce the uniform and staggered moments as
    \begin{align}
        \mqty(
            \vec{\mathcal{M}}_{\mathrm{u}} \\
            \vec{\mathcal{M}}_{\mathrm{s}}
        )
        =
        \frac{1}{\sqrt{2}} \mqty(
            \hat{1} & \hat{1} \\
            \hat{1} & -\hat{1}
        ) \mqty(
            \vec{\mathcal{M}}_{\mathrm{A}} \\
            \vec{\mathcal{M}}_{\mathrm{B}}
        ).
    \end{align}
    This expression is useful in analyzing the mean-field solutions shown later.

Now we explain the method of numerical calculation.
The solutions are obtained by renewing the order parameters iteratively using the self-consistent equation.
The free energy and the self-consistent equation are given by
\begin{align}
    \mathcal F &= -T \ln Z,
    \\
    \vec{\mathcal M}_\al &= - \frac{\partial \mathcal F}{\partial \vec {H}_\al},
\end{align}
where $Z = \Tr \epn^{-\beta \mathcal H_{\rm MF}}$ is the partition function made of the mean-field Hamltonian. For the derivation of the self-consistent equation, the parameters $\vec H$ and $\vec {\mathcal M}$ must be regarded as independent variables.

    The system with the present effective Hamiltonian has 35 kinds of order parameters per site, and there may exist several solutions which take the same free energy as they are connected by symmetries.
    In the next sections, we show the simplest form of the order parameters among those energetically degenerate solutions.

\subsection{Response functions} \label{sec:response}
    In this subsection, we consider the response function to the weak static field.
    We expand the mean-field Hamiltonian up to first order of the field
    \begin{align}
        &\mathcal{H}^{\mathrm{MF}}
        =
        \mathcal{H}^{(0)} + \mathcal{H}^{(1)} + \order{H^2},
        \\
        &\mathcal{H}^{(0)}
        =
        \sum_{i,j} \qty(\vec{\mathcal{M}}_i^{(0)})^{\mathrm{T}} \hat{I}_{ij} \vec{O}_j,
        \\
        &\mathcal{H}^{(1)}
        =
        -\sum_{i,j} \qty[\vec{H}_i^{\mathrm{T}} {\delta}_{ij}  - \qty(\vec{\mathcal{M}}_i^{(1)})^{\mathrm{T}} \hat{I}_{ij}] \vec{O}_j,
    \end{align}
    where the superscript represents the perturbative order of the field and we have neglected the constant term.
    When we define the effective field as $\vec{\tilde{H}}_i = \vec{H}_i - \sum_j \hat{I}_{ji}^{\mathrm{T}} \vec{\mathcal{M}}_j^{(1)}$ and treat $\mathcal{H}^{(1)}$ as perturbation, we obtain the following linear response relation
    \begin{align} \label{eq:linear_response}
        \vec{\mathcal{M}}_i^{(1)}
        =
        \sum_j \hat{\chi}^{(0)}_{ij} \vec{\tilde{H}}_j
        =
        \sum_j \hat{\chi}_{ij} \vec{{H}}_j,
    \end{align}
where $\hat \chi$ is the full susceptibility for the bare external field $\vec H_i$.
According to linear response theory, 
the zeroth-order susceptibility is obtained by
    \begin{align}
        \hat\chi^{(0)}_{ij}
        =
        \int_0^{1/T} \dd{\tau} \qty[
            \ev{T_{\tau} \vec O_i  \vec O_j^{\rm T} \qty(\tau)}_{0}
            -
            \vec{\mathcal M}_i^{(0)} \qty(\vec{\mathcal M}_j^{(0)})^{\rm T}
        ],
        \label{eq:zeroth_chi}
    \end{align}
    where $\tau$ is an imaginary time and $T_{\tau}$ is imaginary time ordering operator.
    The Heisenberg picture in an imaginary time is expressed as
    \begin{align}
        \vec O_i (\tau)
        =
        \epn^{\tau \mathcal{H}^{(0)}} \vec O_i \epn^{-\tau \mathcal{H}^{(0)}}.
    \end{align}
    $\ev{\cdots}_0$ represents the statistical average with $\mathcal{H}^{(0)}$.
    The susceptibility matrix $\hat{\chi}^{(0)}_{ij}$ has only intra-site component since each site is independent under MFA.
    Substituting the concrete expression to the effective field in Eq.~\eqref{eq:linear_response}, we obtain
    \begin{align}
        \sum_j \qty[
            {\delta}_{ij} \hat 1
            +
            \sum_k \hat{{\chi}}^{(0)}_{ik} \hat{I}_{kj}^{\mathrm{T}}
        ] \vec{\mathcal{M}}_j^{\qty(1)}
        =
        \sum_j \hat{{\chi}}^{(0)}_{ij} \vec{H}_j.
    \end{align}
    Then, taking matrix inverse of the left hand side and combining it with Eq.~\eqref{eq:linear_response}, we obtain the susceptibility matrix $\hat \chi_{ij}$.
    For a bipartite lattice, we introduce the uniform and staggered susceptibilities by
    \begin{align}
    \hat \chi_{\mathrm{u}} &= \frac{1}{N} \sum_{i,j} \hat \chi_{ij},
    \label{eq:def_chi_u}
    \\
    \hat \chi_{\mathrm{s}} &= \frac{1}{N} \sum_{i,j} s_i s_j \hat \chi_{ij},
    \label{eq:def_chi_s}
    \end{align}
    where $s_i = +1$ for $i\in {\rm A}$ and $s_i = -1$ for $i\in {\rm B}$.
    This quantity will be shown in the next section.
    Although we focus on the static response functions in this paper, the above argument can easily be generalized for the dynamical susceptibility which captures the magnetic and electric dynamics of the localized model.

    From the view point of Landau theory, we can also discuss the stability of the solution based on the susceptibilities.
    We write down the Landau free energy with an order parameter up to second order as
    \begin{align}
        \mathcal{F}_{\mathrm{L}}
        =
        \frac 1 2 \sum_{i,j} \vec{\mathcal{M}}_i^{\mathrm{T}} \hat{a}_{ij} \vec{\mathcal{M}}_j
        -
        \sum_{i} \vec{H}_i^{\mathrm{T}} \vec{\mathcal{M}}_i,
    \end{align}
    where $\hat{a}_{ij}$ is a coefficient of the quadratic term.
    Note that, here, $\vec {\mathcal M}$ is defined as the deviation from its equilibrium point.
    Then we obtain the following equation of states:
    \begin{align}
        \sum_j \hat{a}_{ij} \vec{\mathcal{M}}_j
        =
        \vec{H}_i.
    \end{align}
    Comparing the linear response function, we find that the Hessian matrix is identical to the inverse susceptibility:
    \begin{align} \label{eq:hessian}
        \pdv{\mathcal{F}_{\mathrm{L}}}{\vec{\mathcal{M}}_i}{\vec{\mathcal{M}}_j}
        =
        \hat{a}_{ij}
        =
        (\hat \chi^{-1})_{ij}.
    \end{align}
    We can consider the necessary and sufficient condition for the stable solution.
    Let $\ep_{n}$ be the $n$-th eigenvalue of the matrix $\hat{a}_{ij}$.
    Each energy corresponds to the eigenenergy of the excitation modes.
    We must have the condition
    \begin{align}
        \ep_{n} \geq 0,
    \end{align}
    for all $n$, if the system is thermodynamically stable.
    If $\ep_{n} = 0$ is obtained, it indicates the presence of the Nambu-Goldstone mode.
    With use of Eq.~\eqref{eq:hessian}, in the actual calculations, we obtain $\ep_{n}$ by diagonalizing the inverse susceptibility matrix.

\section{Numerical results for spherical models} \label{sec:results_spherical}

In the following of this paper, we will encounter the successive phase transitions with decreasing temperature.
There, we denote each transition temperature as $T_{c1}>T_{c2}>\cdots$.
If there is only one transition temperature is identified, we use $T_c$ to denote it.
Note that we use the same symbol for the transition temperatures in different models.

\subsection{Spherical spin-orbital model} \label{sec:effective_model}
First we consider the model in the spherical limit.
    Namely, we assume the hopping matrix given in Eq.~\eqref{eq:H_t} as
    \begin{align}
        \hat t_{ij}
        =
        \mqty(
            t & 0 & 0 \\
            0 & t & 0 \\
            0 & 0 & t
        ),
    \end{align}
    for a bipartite lattice with the coordination number $z$.
    Using the spin-orbital operator $O_i^{\eta\mu}$ defined in the previous section, we obtain the spherical model as
    \begin{align} \label{eq:H_eff_spherical}
        &\mathcal{H}_\mathrm{eff}
        =
        -\sum_{\la ij \ra}\Big[
        I_S \bm{S}_i \cdot \bm{S}_j
        +
        I_L \bm{L}_i \cdot \bm{L}_j
        +
        I_Q \sum_{\eta} Q_i^{\eta} Q_j^{\eta}
        \nonumber
        \\
        &+
        I_R \sum_{\mu} \sum_{\nu} R_i^{\nu,\mu} R_j^{\nu,\mu}
        +
        I_T \sum_{\mu} \sum_{\eta} T_i^{\eta,\mu} T_j^{\eta,\mu}
        +
        I_0\Big],
    \end{align}
    where the sum with $\la ij \ra$ is taken over the pairs of the NN sites.
The superscript $\mu, \nu$ ($=x,y,z$) and $\eta$ ($=x^2-y^2,z^2,xy,yz,zx$) are the indices for the polynomials, which represents the component of the spin, rank 1 orbital and rank 2 orbital, respectively.
    We have rewritten the operators in accordance with their symmetries as
    \begin{align}
        &S_i^{\mu}
        =
        \frac{1}{2} O_i^{0\mu},
        \label{eq:def_operator_S}
        \\
        &L_i^{x}
        =
        \frac{1}{2} O_i^{70},\ 
	    L_i^{y}
        =
        \frac{1}{2} O_i^{50},\ 
	    L_i^{z}
        =
        \frac{1}{2} O_i^{20},
        \label{eq:def_operator_L}
        \\
        &Q_i^{x^2-y^2}
        =
        \frac{1}{2} O_i^{30},\ 
        Q_i^{z^2}
        =
        \frac{1}{2} O_i^{80},
        \nonumber
        \\
        &Q_i^{xy}
        =
        \frac{1}{2} O_i^{10},\ 
        Q_i^{yz}
        =
        \frac{1}{2} O_i^{60},\ 
        Q_i^{zx}
        =
        \frac{1}{2} O_i^{40},
        \label{eq:def_operator_Q}
        \\
        &R_i^{x,\mu}
        =
        \frac{1}{2} O_i^{7\mu},\ 
        R_i^{y,\mu}
        =
        \frac{1}{2} O_i^{5\mu},\ 
        R_i^{z ,\mu}
        =
        \frac{1}{2} O_i^{2\mu},\ 
        \label{eq:def_operator_R}
        \\
        &T_i^{x^2-y^2,\mu}
        =
        \frac{1}{2} O_i^{3\mu},\ 
        T_i^{z^2,\mu}
        =
        \frac{1}{2} O_i^{8\mu},
        \nonumber
        \\
        &T_i^{xy,\mu}
        =
        \frac{1}{2} O_i^{1\mu},\ 
        T_i^{yz,\mu}
        =
        \frac{1}{2} O_i^{6\mu},\ 
        T_i^{zx,\mu}
        =
        \frac{1}{2} O_i^{4\mu}.
        \label{eq:def_operator_T}
    \end{align}
    The physical meaning of each order parameter now becomes clearer with this notation.
    We call $S_i^{\mu}$ a magnetic spin (MS or $S$), $L_i^{\mu}$ a magnetic orbital (MO or $L$), $Q_i^{\mu}$ a electric orbital (EO or $Q$), $R_i^{\nu,\mu}$ a electric spin-orbital (ESO or $R$) and $T_i^{\eta,\mu}$ a magnetic spin-orbital (MSO or $T$) moments.
    $I_0$ represents energy gain by the second order perturbation process.
    Obviously, Eq.~\eqref{eq:H_eff_spherical} satisfies SU(2)$\times$SO(3) symmetry in spin-orbital space.

    We will show the numerical results of the $n = 1$ and $n = 3$ spherical models under MFA, both of which have the six states per site in the model space as discussed in Sec.~\ref{sec:hubbard}.
    We beforehand introduce the following notation with regard to the coupling constants defined in Eq.~\eqref{eq:H_eff_spherical} as
    \begin{align} \label{eq:coupling_constants_decomp}
        I_{\xi} &= \sum_{n} A_{\xi n} \frac{t^2 }{\Delta E_n},
    \end{align}
    for $\xi = S,L,Q,R,T,0$, where $\Delta E_n$ represents all possible excitation energies.
    Its energy corresponds to the denominator of Eq.~\eqref{eq:H_eff}.
    The coefficient $A$ is summarized in the tables in the following subsections (see Sec.~\ref{sec:results_n=1} or Sec.~\ref{sec:results_n=3}).

Before we show the mean-field results, we discuss the ground state wave function for the two-site problem.
Using the single site state defined in Eq.~\eqref{eq:ground_state_single_site} or \eqref{eq:wf_n=1}, we obtain the two-site (i.e., sites at $i$ and $j$) ground state as
    \begin{align}
        \ket{\mathrm{gs}}
        =
        \sum_{\gm_i, \sg_i} \sum_{\gm_j, \sg_j} C_{\gm_i \sg_i, \gm_j \sg_j} \ket{\gm_i, \sg_i}_i \ket{\gm_j, \sg_j}_j.
    \end{align}
    The explicit form of the matrix $C$ is written as
    \begin{align}
        \hat{C}
        =
        \lambda^0 \otimes \qty(-\imu \sg^y).
    \end{align}
    This shows that the ground-state wave function is spin-singlet and 
    symmetric on the orbital.
    This is valid for all the spherical cases considered in this section.
    For an infinite lattice, as in the single-orbital Hubbard model, the inter-site spin-singlet state may favor the antiferromagnetic state in the ground state for a bipartite lattice.

\subsection{$n = 1$ model} \label{sec:results_n=1}

First of all, we consider the results for the $n=1$ model.
Although the results are not relevant to the alkali-doped fullerides, the knowledge is useful in interpreting the more complicated model for the spherical $n=3$ model (Sec.~\ref{sec:results_n=3}), the realistic A15- (Sec.~\ref{sec:results_a15}) and fcc-structure fullerides (Sec.~\ref{sec:results_fcc}).

\subsubsection{Coupling constant}

    \begin{table}[t]
        \centering
        \caption{
            Coefficients $A$ defined in Eq.~\eqref{eq:coupling_constants_decomp} for $n = 1$ spherical model.
            The ground state energy is zero.
            We add the details for the intermediate state in the main text.
        }
        \begin{tabular}{cccc} \hline
            $\Delta E_n$ &\ $U - 3J$\ &\ $U - J$\ &\ $U + 2J$\ \\ \hline
            $I_S$ & $-2$ & $10/3$ & $2/3$ \\
            $I_L$ & $3$ & $-5/3$ & $2/3$ \\
            $I_Q$ & $3$ & $-1/3$ & $-2/3$ \\
            $I_R$ & $1$ & $5/3$ & $-2/3$ \\
            $I_T$ & $1$ & $1/3$ & $2/3$ \\
            $I_0$ & $-6$ & $-10/3$ & $-2/3$ \\ \hline
        \end{tabular}
        \label{tab:coupling_constants_n=1}
    \end{table}
    We begin with the analysis of the intermediate states relevant to the second-order perturbation theory.
    We show the coefficients $A$ defined in Eq.~\eqref{eq:coupling_constants_decomp} in Table~\ref{tab:coupling_constants_n=1}.
    We have the three kinds of excited states, whose energy is determined by the local Coulomb interaction.
    For $\Delta E_n = U-3J$, the intermediate states are nine-fold degenerate spin-triplet states, as expressed, e.g., by $c_{i,y,\ua}^{\dg}c_{i,x,\ua}^{\dg} \Ket{0}$ and $\frac{1}{\sqrt{2}}\qty(c_{i,y,\da}^{\dg}c_{i,x,\ua}^{\dg} + c_{i,y,\ua}^{\dg}c_{i,x,\da}^{\dg}) \Ket{0}$.
    For $\Delta E_n = U-J$, the intermediate states are the inter-orbital spin-singlet states such as $\frac{1}{\sqrt{2}} \qty(c_{i,y,\da}^{\dg}c_{i,x,\ua}^{\dg} - c_{i,y,\ua}^{\dg}c_{i,x,\da}^{\dg}) \Ket{0}$, and the intra-orbital spin-singlet states with anti-bonding orbitals written as $\frac{\sqrt{2}}{3} \qty(2b_{i,z}^{\dg} - b_{i,x}^{\dg} - b_{i,y}^{\dg}) \Ket{0}$.
    These two kinds of states take the same energy since there is the spherically symmetric condition $U' = U - 2J$.
    For $\Delta E_n = U+2J$, there is only one intermediate state, which is intra-orbital spin singlet and bonding state written as $\frac{1}{\sqrt{3}} \qty(b_{i,x}^{\dg} + b_{i,y}^{\dg} + b_{i,z}^{\dg}) \Ket{0}$.

    \begin{figure}[t]
    \centering
    \includegraphics[width=85mm]{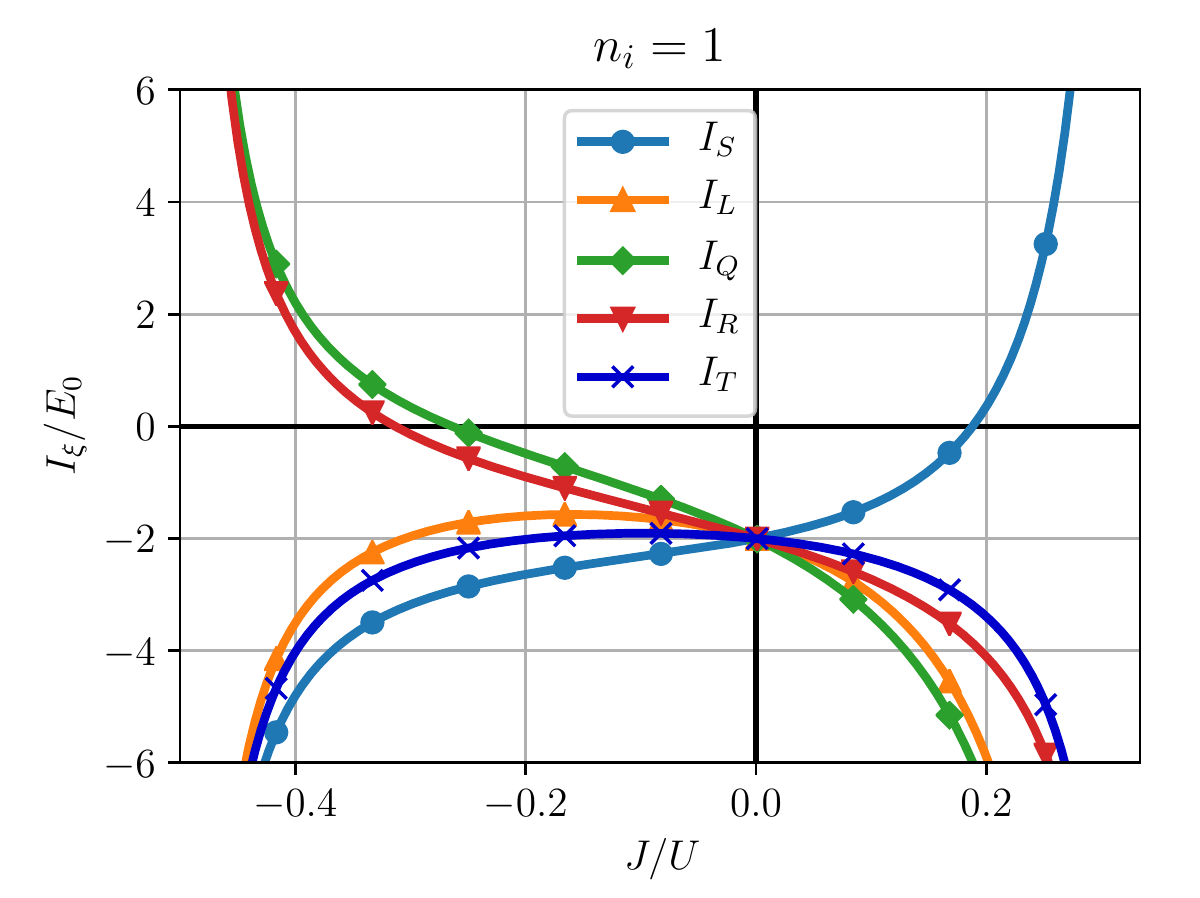}
    \caption{
        Hund's coupling ratio $J/U$ dependence of the coupling constants for $n = 1$ spherical model.
        The vertical axis is normalized by $E_0 = t^2/U$.
    }
    \label{fig:coupling_constants_n=1}
    \end{figure}
    We show the Hund's coupling dependence of the coupling constants in Fig.~\ref{fig:coupling_constants_n=1}.
    The perturbation theory is justified for $-1/2 < J/U < 1/3$ where the ground states are written in the form of Eq.~\eqref{eq:wf_n=1}.
    Taking $J = 0$, the coupling constants become identical.
    This reflects that the system has SU(6) symmetry and the degrees of freedom of the spin and orbital are equivalent in the absence of Hund's coupling.
    The largest coupling constant is $I_S$ for the antiferromagnetic case ($J < 0$) and $I_Q$ for the ferromagnetic Hund's coupling ($J > 0$).
    This shows that the system tends to be antiferromagnetic (AFM) or antiferro-orbital (AFO) order depending on the sign of the Hund's coupling.
    This is understood from the intermediate state.

    In the case of $J > 0$,  which is relevant to the usual $t_{2g}$-orbital $d$-electron systems with $n=1$ per atom, the energetically favorable intermediate two-electron state is inter-orbital spin triplet.
    To realize this intermediate state, the initial state needs to occupy parallel spin configuration with different orbitals such as $c_{i,x,\ua}^{\dg}c_{j,y,\ua}^{\dg} \Ket{0}$.
    Therefore, the orbital order should be dominant for $J > 0$ as a leading-order ordering instability.
    If we take $J/U \gtrsim 0.2$, $I_S$ takes a ferromagnetic coupling constant, which favors parallel spins at two sites.

    As for $J < 0$, on the other hand, the intermediate state tends to be intra-orbital spin singlet and bonding state.
    The corresponding initial state must be antiparallel spin with the same orbital such as $c_{i,x,\ua}^{\dg}c_{j,x,\da}^{\dg} \Ket{0}$.
    Thus, the magnetic order should be dominant for $J < 0$.

\subsubsection{Mean-field solutions for antiferromagnetic Hund's coupling ($J<0$)}

    \begin{figure}[t]
    \centering
	\includegraphics[width=85mm]{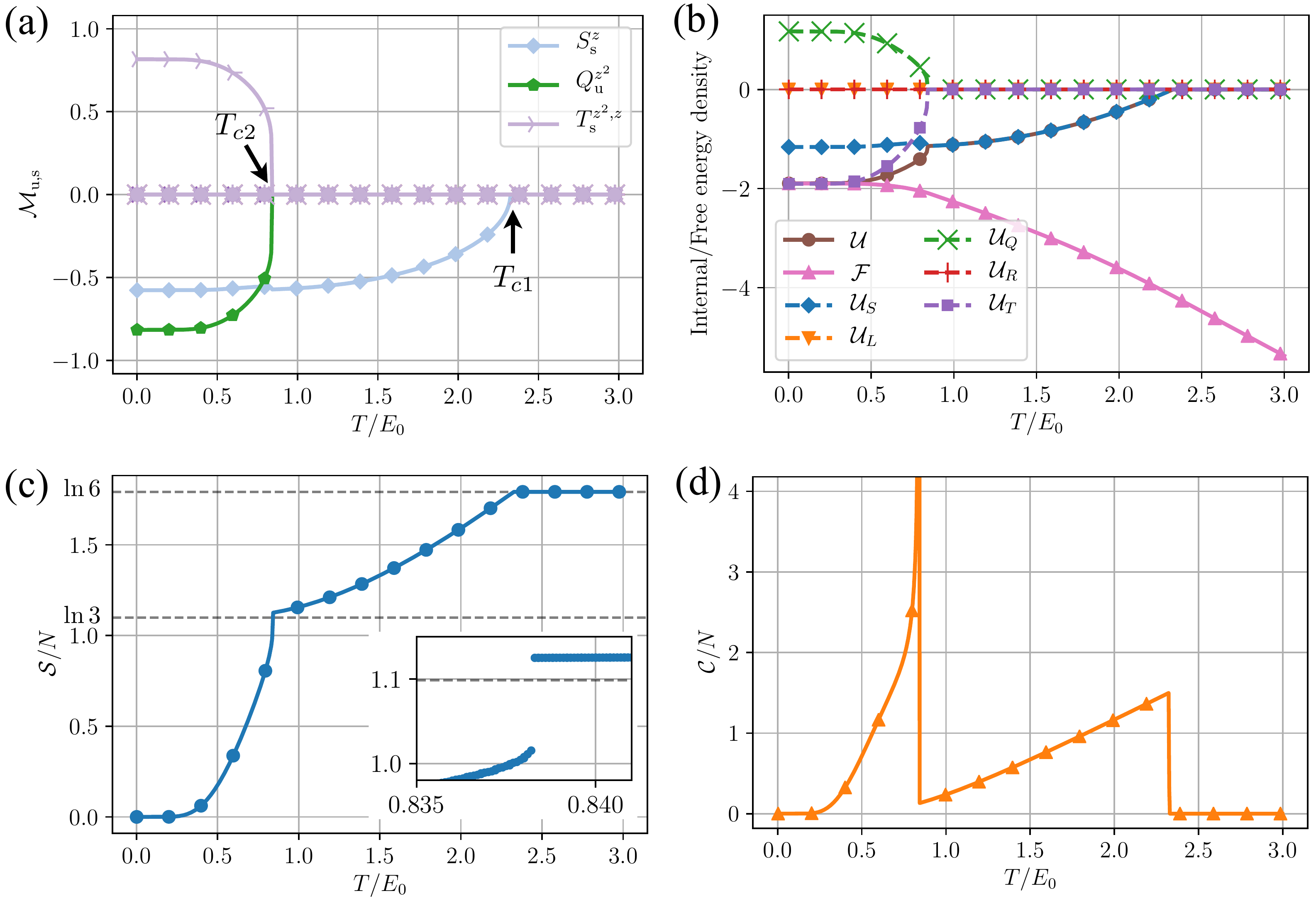}
    \caption{
		Temperature dependence of (a) the order parameter, (b) the decomposed internal energy and total free energy density, (c) entropy and (d) specific heat for $n = 1, J/U = -0.1$ bipartite spherical model.
		The inset in (c) is enlarged plot around $T_{c2}$.
		The energy unit of these plots are $E_0 = t^2/U$.
	}
	\label{fig:temperature_n=1_spherical_J=-0.1_stagg}
    \end{figure}
    
	Let us turn our attention to the numerical results using MFA in the spherical model.
	We take the NN coordination number $z=6$ in the numerical calculation by assuming a simple cubic lattice in three dimensions.
	Figure~\ref{fig:temperature_n=1_spherical_J=-0.1_stagg} shows the temperature dependence of the physical quantities in the bipartite lattice model at $J/U = -0.1$ (antiferromagnetic Hund's coupling).
	We take $E_0 \equiv t^2/U$ as the unit of energy.
The uniform and staggered order parameters are shown in Fig.~\ref{fig:temperature_n=1_spherical_J=-0.1_stagg}(a), where the antiferromagnetic spin (AF-$S$) order appears first with decreasing temperature from the high-temperature limit.
This corresponds to the largest coupling constant $I_S$ in Fig.~\ref{fig:coupling_constants_n=1}.
At lower temperatures, the ferro (F)-orbital $Q$ moment of $z^2$ type appears together with
the AF-$T$ (MSO) moments.
In order to clarify which is the primary order parameter of the second phase transition at $T_{c2}$, we show in Fig.~\ref{fig:temperature_n=1_spherical_J=-0.1_stagg}(b) the internal energy
and free energy per site, where the internal energy is decomposed into each contribution as
	\begin{align} \label{eq:internal_ene_decomp}
    &\mathcal U_S = I_S \la \bm S_{\rm A} \ra \cdot \la \bm S_{\rm B} \ra,
    \\
    &\mathcal{U}_L = I_L \la \bm L_{\mathrm{A}} \ra \cdot \la \bm L_{\mathrm{B}} \ra,
    \\
    &\mathcal{U}_Q = I_Q \sum_{\eta} \la Q_{\mathrm{A}}^{\eta} \ra \la Q_{\mathrm{B}}^{\eta} \ra,
    \\
    &\mathcal{U}_R = I_R \sum_{\mu} \sum_{\nu} \la R_{\mathrm{A}}^{\nu,\mu} \ra \la R_{\mathrm{B}}^{\nu,\mu} \ra,
    \\
    &\mathcal{U}_T = I_T \sum_{\mu} \sum_{\eta} \la T_{\mathrm{A}}^{\eta,\mu} \ra \la T_{\mathrm{B}}^{\eta,\mu} \ra.
	\end{align}
The total internal energy is given by $\mathcal U = \sum_\xi \mathcal U_\xi$ for $\xi=S,L,Q,R,T$, where the energy is measured from $I_0$.
We see from Fig.~\ref{fig:temperature_n=1_spherical_J=-0.1_stagg}(b) that the energy $\mathcal U_T$ is gained below $T_{c2}$ but $\mathcal U_Q$ is not.
Hence, the AF-$T$ should be the primary order parameter and F-$Q$ is just induced by the combination of AF-$S$ plus AF-$T$ moments.
The results are consistent with the magnitude relation $I_T > I_Q$ seen in Fig.~\ref{fig:coupling_constants_n=1}, where the larger energy gain is obtained from $T$-moment than the energy loss from $Q$.

Figure~\ref{fig:temperature_n=1_spherical_J=-0.1_stagg}(c) shows the temperature dependence of the entropy, where all the entropy is released in the ground state.
With increasing temperature, the entropy shows a kink at $T/E_0 \simeq 0.84$, at which the value of the entropy is close to $\ln 3$ meaning that the orbital degeneracy is lifted below this transition temperature.
The inset of (c) shows the magnified picture of the entropy near $T_{c2}$, indicating the first-order transition.
The specific heat $\mathcal{C} =\partial \mathcal{U} / \partial T$ is also shown in Fig.~\ref{fig:temperature_n=1_spherical_J=-0.1_stagg}(d).
There are two discontinuity corresponding to the spin and orbital orders.

    \begin{figure}
        \centering
        \includegraphics[width=85mm]{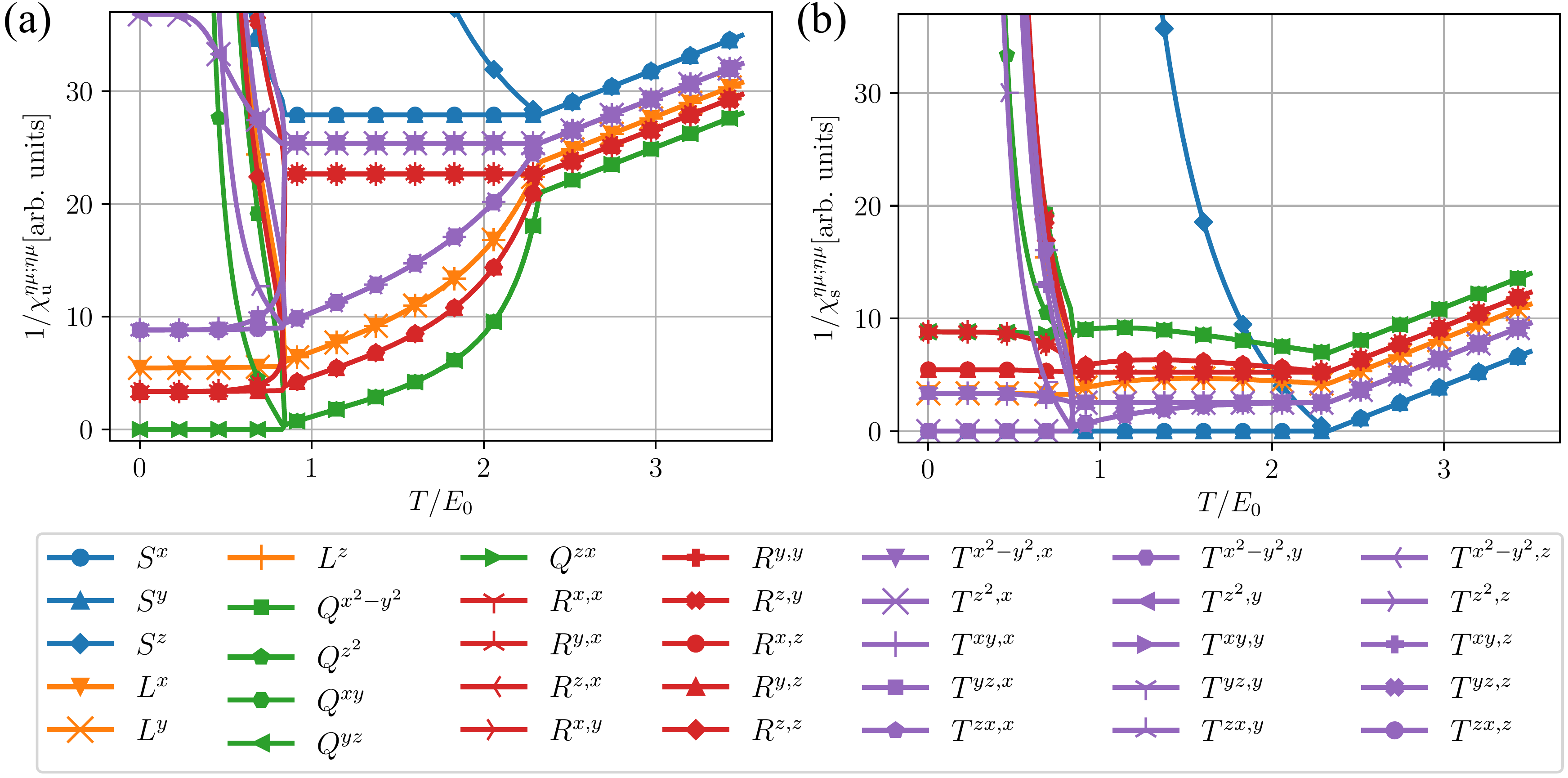}
        \caption{
        Temperature dependence of the inverse of (a) uniform and (b) staggered component of the diagonal susceptibilities.
        The energy unit is $E_0 = t^2/U$.
        }
        \label{fig:temperature_n=1_spherical_J=-0.1_stagg_chi}
    \end{figure}

Next we show in Fig.~\ref{fig:temperature_n=1_spherical_J=-0.1_stagg_chi} the inverse of the diagonal susceptibilities $\chi_{\mathrm{u}}^{\eta\mu;\eta\mu}$ (uniform) and $\chi_{\mathrm{s}}^{\eta\mu;\eta\mu}$ (staggered) which are defined in Eqs.~\eqref{eq:def_chi_u} and \eqref{eq:def_chi_s}.
First, we observe that the susceptibilities shown here are all positive, indicating a stable solution.
The AF-$S$ susceptibility of $x,y,z$ type diverges at $T/E_0 \simeq 2.3$ signaling the onset of the antiferromagnetic order.
Below this transition temperature, the longitudinal $z$ component is decreased while the perpendicular $x,y$ components remain divergent.
This behavior indicates the presence of the Goldstone mode, where the excitations are induced by rotating the $z$ component into
$xy$-plane, as in the standard Heisenberg model.
Inside this magnetic phase, the orbital (F-$Q$) and spin-orbital (AF-$T$) susceptibility, which are $z^2$ type in orbital part, continue to grow and tend to diverge at lower transition point ($T_{c2}$).
As shown in Fig.~\ref{fig:temperature_n=1_spherical_J=-0.1_stagg_chi}(a),
the `perpendicular' components, i.e. F-$Q^{yz}$, F-$Q^{zx}$, remain divergent below $T_{c2}$, indicating the presence of the Goldstone mode even for the orbital order in the spherical model.
Namely, because of the symmetry of the spin-orbital space, the energetically equivalent solutions exist and are obtained by rotating the order parameters.

Next we discuss the ground state wave function, which includes the information of order parameter at zero temperature limit.
As is evident from the zero entropy at $T=0$, we have the non-degenerate ground state.
In the present case, the ground state wave function is very simple and is given using 
Eq.~\eqref{eq:ground_state_single_site}
by
	\begin{align}
		&\Ket{\psi_{\mathrm{A}}}
		=
		\Ket{n=1, z,\da}_{\mathrm{A}},
		\\
		&\Ket{\psi_{\mathrm{B}}}
		=
		\Ket{n=1, z,\ua}_{\mathrm{B}},
	\end{align}
for each sublattice.
This corresponds to the staggered spin ordered and uniform orbital ordered state, as is consistent with Fig.~\ref{fig:temperature_n=1_spherical_J=-0.1_stagg}(a).
More specifically, we can construct the order parameters from the direct product of the wave functions.
In the present case, we obtain at sublattice $\al$ as
\begin{align}
\Ket{\psi_\al} \Bra{\psi_\al}
&= \mp\frac{1}{\sqrt{6}} S^z_\al -\frac{1}{\sqrt{3}} Q^{z^2}_\al \pm \frac{1}{\sqrt{3}} T^{z^2,z}_\al + \frac{1}{6},
\end{align}
where the operators are defined in Eqs.~\eqref{eq:def_operator_S}--\eqref{eq:def_operator_T}.
The upper (lower) sign is chosen for $\al = \mathrm{A}$ ($\al = \mathrm{B}$).
The quantities that appear in the right-hand side are identical to the order parameters shown in Fig.~\ref{fig:temperature_n=1_spherical_J=-0.1_stagg}(a).

\subsubsection{Mean-field solutions for ferromagnetic Hund's coupling ($J>0$)}

    \begin{figure}[t]
    \centering
	\includegraphics[width=85mm]{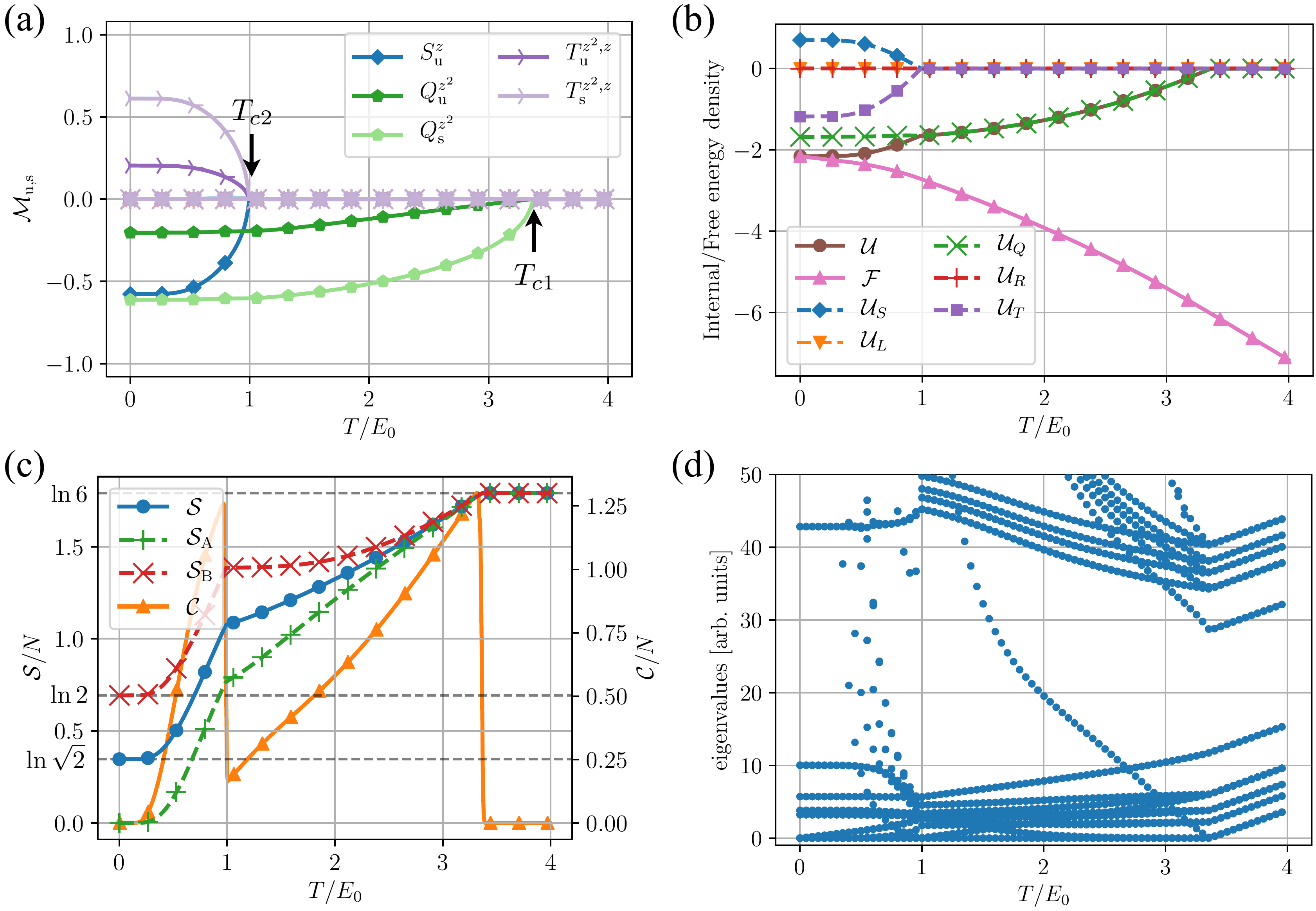}
    \caption{
        Temperature dependence of (a) the order parameter, (b) the decomposed internal energy and total free energy density, (c, left axis) the single site entropy, (c, right axis) the specific heat and (d) the eigenvalues of the Hessian matrix $\hat{a}$ for bipartite spherical model with $n = 1, J/U = 0.1$.
    }
	\label{fig:temperature_n=1_spherical_J=0.1_stagg}
	\end{figure}

We show the results for the $J/U=0.1$ case, where the model is now relevant to materials with $d$-electrons, to make contrast with behaviors of the systems with antiferromagnetic Hund's coupling.
Figure~\ref{fig:temperature_n=1_spherical_J=0.1_stagg}(a) shows the temperature evolution of the order parameters.
As seen in Fig.~\ref{fig:coupling_constants_n=1}, the largest coupling constant is $I_Q$ which is antiferro ($I_Q<0$), and therefore the AF-$Q$ order of $z^2$-type appears at the highest transition temperature ($T_{c1}$).
The F-$Q$ order of the same $z^2$-type is simultaneously induced.
The rise of the order parameters near the transition temperature behaves as $\sim \sqrt{T_{c1}-T}$ for AF-$Q$ and $\sim T_{c1}-T$ for F-$Q$.
Hence the AF-$Q$ is the primary order.
From the symmetry argument, it can be shown that the F-$Q$ order arises from AF-$Q$ order since the coupling term in the Landau free energy has the form $Q^{z^2}_{\rm u} (Q^{z^2}_{\rm s})^2$.
The existence of such third-order term can be understood if one considers the symmetry in the plane of $Q^{z^2}$-$Q^{x^2-y^2}$~\cite{Hoshino2017}.
At lower temperatures, the magnetic F-$S$ order appears, where $T$-moments of $T^{z^2,z}$-type are also finite.
From the internal-energy analysis shown in Fig.~\ref{fig:temperature_n=1_spherical_J=0.1_stagg}(b), the relevant ordering at $T_{c2}$ is induced from the interaction $I_T$ while $I_S$ is energetically unfavorable.
Thus, comparing with the $J/U=-0.1$ case, the roles of magnetic order and electric (orbital) order are switched.
This switching of the magnetic and orbital ordered states depending on the sign of $J$ has also been reported in the two orbital model~\cite{Steiner2016}.

We next show the temperature dependence of the entropy and specific heat in Fig.~\ref{fig:temperature_n=1_spherical_J=0.1_stagg}(c), where we have defined the sublattice-dependent entropy (Shannon entropy) by
	\begin{align} \label{eq:shannon_ent}
		\mathcal{S}_{\al}
		=
		- \sum_n p_n^{\al} \ln p_n^{\al},
	\end{align}
where $p_n^\al$ is the probability for the $n$-th state as calculated from the local partition function $Z_{\al} = \sum_n \exp(-\beta E_{n}^\al ) = \sum_n p_{n}^\al Z_\al$.
Since the entropy at zero temperature is zero at A sublattice and is finite at B sublattice, the two sublattices are inequivalent and are not simply connected by symmetry operations.
This is due to the presence of the both uniform and staggered orbital order parameters in Fig.~\ref{fig:temperature_n=1_spherical_J=0.1_stagg}(a).
Indeed, the wave function in the ground state is written for each sublattice as
	\begin{align}
		&\Ket{\psi_{\mathrm{A}}}
		=
		\Ket{n=1,z,\da}_{\mathrm{A}},
		\\
		&\Ket{\vec{\psi}_{\mathrm{B}}}
		=
		\mqty(
			\Ket{n=1,x,\da}_{\mathrm{B}} \\
			\Ket{n=1,y,\da}_{\mathrm{B}}
		).
	\end{align}
The remaining degeneracy at B sublattice is because the $x$- and $y$-orbital components are equivalent.
Namely, the triply degenerate state at each sublattice splits depending on the sublattice: $z$ orbital becomes energetically higher at A sublattice and lower at B sublattice.
Thus the antiferro order of this type cannot lift the degeneracy completely.

Usually, the degeneracy is lifted by the interaction effects and the unique ground state is expected.
Then, one may suspect that the remaining degeneracy might indicate the instability of the solutions.
In order to show that our degenerate ground states are really stable, we show the energy spectra of the Hessian matrix discussed in Sec.~\ref{sec:response}. 
As shown in Fig.~\ref{fig:temperature_n=1_spherical_J=0.1_stagg}(d), the excitation energies in terms of Landau theory are all positive or zero, and the system is thus stable.
The degeneracy at $T=0$ is due to the absence of the relevant interactions, and will be resolved once the other types of the interaction are included in the more realistic situations.

    We comment on the case where we allow only for the uniform solutions, by having the geometrical frustration effect in mind which does not favor a simple staggered orders.
    Actually, the $n = 1$ uniform spherical model around $J = 0$ has no solution at any temperature because all of the coupling constants are negative (antiferromagnetic) in the spherical model (see Fig.~\ref{fig:coupling_constants_n=1}).
    On the other hand, for relatively large $|J|$ region the uniform solutions can exist.
    However, since the typical value of Hund's coupling is $\abs{J}/U \sim 0.1$ or less, we do not enter the regime with larger $\abs{J}$ in this paper.

\subsection{$n = 3$ model} \label{sec:results_n=3}

Here we consider the model with three electrons per molecule and with the antiferromagnetic Hund's coupling ($J<0$).
This model is more relevant to the existing fullerides with half-filled $t_{1u}$ molecular orbitals.

\subsubsection{Coupling constants}

    \begin{table}[t]
        \centering
        \caption{
            Coefficients $A$ in Eq.~\eqref{eq:coupling_constants_decomp} for $n = 3$ spherical model.
            The ground state is written as $\Ket{\gm_i, \sg_i}_i \Ket{\gm_j, \sg_j}_j$ and its energy is $2\qty(3U - 4J)$.
            We add the details for the intermediate state in the main text.
        }
        \begin{tabular}{ccccccc} \hline
            $\Delta E_n$ & $U - 8J$ & $U - 6J$ & $U - 4J$ & $U - 3J$ & $U - J$ & $U + 2J$ \\ \hline
            $I_S$ & $1/2$ & $-5/3$ & $25/18$ & $-4/3$ & $20/9$ & $8/9$ \\
            $I_L$ & $9/8$ & $-5/4$ & $25/72$ & $2$ & $-10/9$ & $8/9$ \\
            $I_Q$ & $-9/8$ & $1/4$ & $-1/72$ & $2$ & $-2/9$ & $-8/9$ \\
            $I_R$ & $-1/8$ & $-5/12$ & $-25/72$ & $2/3$ & $10/9$ & $-8/9$ \\
            $I_T$ & $1/8$ & $1/12$ & $1/72$ & $2/3$ & $2/9$ & $8/9$ \\
            $I_0$ & $-9/2$ & $-5$ & $-25/18$ & $-4$ & $-20/9$ & $-8/9$ \\ \hline
        \end{tabular}
        \label{tab:coupling_constants_n=3}
    \end{table}

    We show the coefficients $A$, which is defined by Eq.~\eqref{eq:coupling_constants_decomp}, in Table~\ref{tab:coupling_constants_n=3}.
    Since we consider the half-filled model, the initial and intermediate states for the two-site problem at the sites $i$ and $j$ relevant to $I_{ij}$ are $(n_i,n_j) = (3,3)$ and $\qty(n_i, n_j) = (2, 4)$, respectively.
    Here, $n_i=2$ and $n_i=4$ states are connected with each other by the particle-hole (PH) transformation.
    The explicit form for $n_i = 2$ state is same as those given in Sec.~\ref{sec:results_n=1}, and thereby the $n=4$ can also be constructed from $n=2$ accordingly.
    Below, we list the types of the intermediate states and their energies, specifically focusing on the $n_j=4$ state.

    The intermediate states with the excited energy$\Delta E_n = U - 8J$ are nine kinds of inter-orbital spin triplet state for $n_i = 2$ and the PH transformed states for $n_j = 4$ such as $b_{j,z}^{\dg}c_{j,y,\ua}^{\dg}c_{j,x,\ua}^{\dg} \Ket{0}$.
    For $\Delta E_n = U - 6J$, the intermediate states are the inter-orbital spin triplet states for $n_i=2$ and the PH transformed states which have inter-orbital spin singlet states such as $\frac{1}{\sqrt{2}} b_{j,z}^{\dg} \qty(c_{j,y,\da}^{\dg}c_{j,x,\ua}^{\dg} - c_{j,y,\ua}^{\dg}c_{j,x,\da}^{\dg}) \Ket{0}$ or intra-orbital spin singlet with anti-bonding such as $\frac{\sqrt{2}}{3} \qty(2b_{j,z}^{\dg}b_{j,y}^{\dg} - b_{j,z}^{\dg}b_{j,x}^{\dg} - b_{j,y}^{\dg}b_{j,x}^{\dg}) \Ket{0}$.
    For $\Delta E_n = U - 4J$, the intermediate states are the inter-orbital spin singlet or intra-orbital spin singlet with anti-bonding states for $n_i=2$, and their PH transformed versions for the $j$ site.
    For $\Delta E_n = U - 3J$, the intermediate states are the intra-orbital spin singlet and bonding states for $n_i=2$, and the states which have inter-orbital spin triplet for $n_j=4$.
    For $\Delta E_n = U - J$, the intermediate states are the inter-orbital spin singlet or intra-orbital spin singlet with anti-bonding states ($n_i=2$), and intra-orbital spin singlet and bonding state such as $\frac{1}{\sqrt{3}} \qty(b_{j,z}^{\dg}b_{j,y}^{\dg} + b_{j,z}^{\dg}b_{j,x}^{\dg} + b_{j,y}^{\dg}b_{j,x}^{\dg}) \Ket{0}$ for $n_j=4$.
    Finally, for $\Delta E_n = U + 2J$, which is the lowest among the excited states for $J<0$, the intermediate state is non-degenerate and is written as the intra-orbital spin singlet with bonding state for $n_i=2$ and its PH transformed states for $n_j=4$.

    \begin{figure}[t]
    \centering
    \includegraphics[width=85mm]{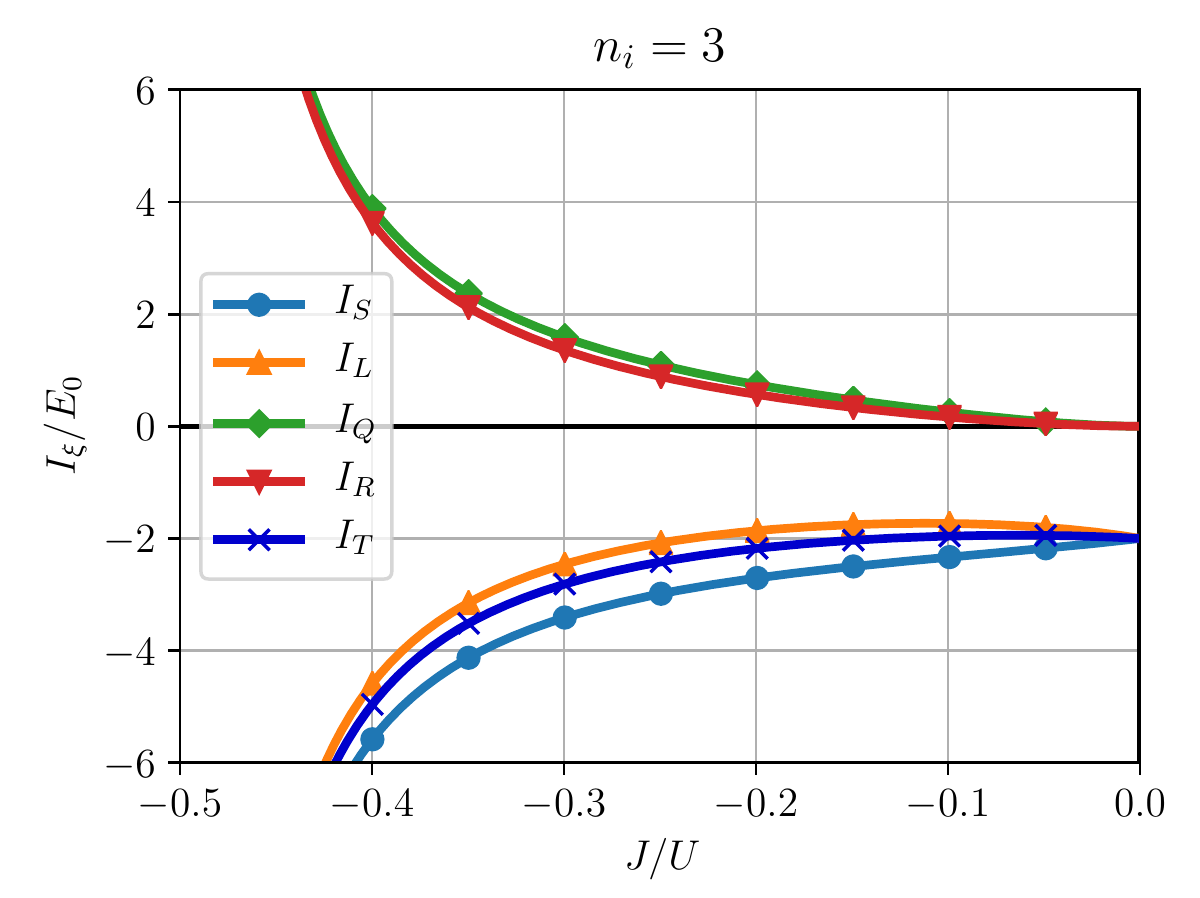}
    \caption{
        Hund's coupling ratio $J/U$ dependence of the coupling constants for $n = 3$ spherical model.
    }
    \label{fig:coupling_constants_n=3}
    \end{figure}
    Figure~\ref{fig:coupling_constants_n=3} shows the Hund's coupling dependence of the coupling constants.
    The perturbation theory is justified for $-1/2 < J/U < 0$ where any level cross for the unperturbed Hamiltonian does not occur.
If we consider $J>0$, the ground state is a total spin $S=3/2$ state (e.g., $c_{i, z, \ua}^\dg c_{i, y, \ua}^\dg c_{i, x, \ua}^\dg |0\ra_i$) and is different from $J<0$.
This point is in contrast with $n=1$ case where the ground state of the local Hamiltonian is not dependent on the sign of $J$ as shown in Fig.~\ref{fig:coupling_constants_n=1}.
It is notable that the coupling constants for $n=3$ case are similar to those of the $n = 1$ spherical model in the region near $J/U = -0.5$, where the same physical behavior is expected.

\subsubsection{Mean-field solutions for bipartite lattice}
    
    \begin{figure*}[t]
    \centering
	\includegraphics[width=140mm]{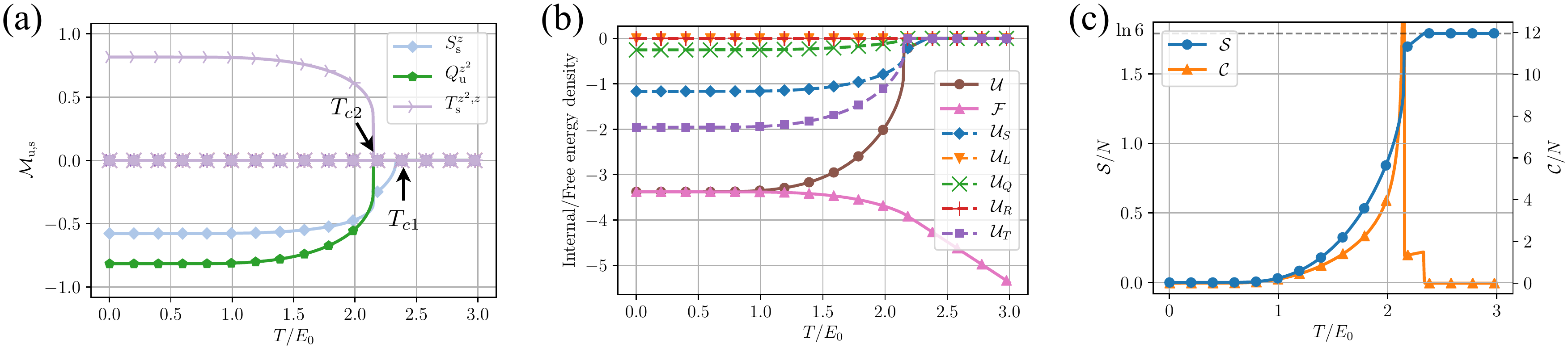}
    \caption{
        Temperature dependence of (a) the order parameter, (b) the decomposed internal energy and total free energy density, (c, left axis) entropy and (c, right axis) the specific heat for $n = 3$ bipartite spherical model with $J/U=-0.1$.
        The horizontal axis are normalized by $E_0 = t^2/U$.
    }
	\label{fig:temperature_n=3_spherical_J=-0.1_stagg}
	\end{figure*}
	
We show in Fig.~\ref{fig:temperature_n=3_spherical_J=-0.1_stagg}(a) the order parameters for the bipartite lattice model with $n=3$ and $J/U=-0.1$.
At $T_{c1} \simeq 2.3E_0$, the system shows the antiferromagnetic order, which is consistent with the largest coupling constant shown in Fig.~\ref{fig:coupling_constants_n=3}.
With decreasing temperature, the second order at $T_{c2}$ appears, where the F-$Q^{z^2}$ and AF-$T^{z^2,z}$ order parameters are additionally induced.
We emphasize that this orbital order is not of the ordinary orbital moment of electrons, but of the doublons relevant to the antiferromagnetic Hund's coupling as discussed in Sec.~\ref{sec:hubbard}.

Figure~\ref{fig:temperature_n=3_spherical_J=-0.1_stagg}(b) shows the temperature dependences of the internal energies and free energy.
We show the order-parameter-resolved energies and all the components decrease upon entering the ordered phase.
While this is in contrast to $n=1$ cases shown in the previous subsections, the largest energy gain arises from the AF-$T$ order.

We show in Fig.~\ref{fig:temperature_n=3_spherical_J=-0.1_stagg}(c) the entropy and specific heat.
The clear jump in the specific heat at $T_{c1}$ indicates the second-order phase transition, and the jump in the entropy at $T_{c2}$ is the fingerprint of the first-order phase transition.
The wave function in the ground state is
	\begin{align}
		\Ket{\psi_{\mathrm{A}}}
		=
		\Ket{z, \da}_{\mathrm{A}},
		\\
		\Ket{\psi_{\mathrm{B}}}
		=
		\Ket{z, \ua}_{\mathrm{B}}.
	\end{align}
The ground state is thus non-degenerate as is consistent with the zero entropy at $T=0$.

\subsubsection{Single-sublattice solution} \label{sec:results_n=3_unif}

    \begin{figure*}[t]
    \centering
	\includegraphics[width=140mm]{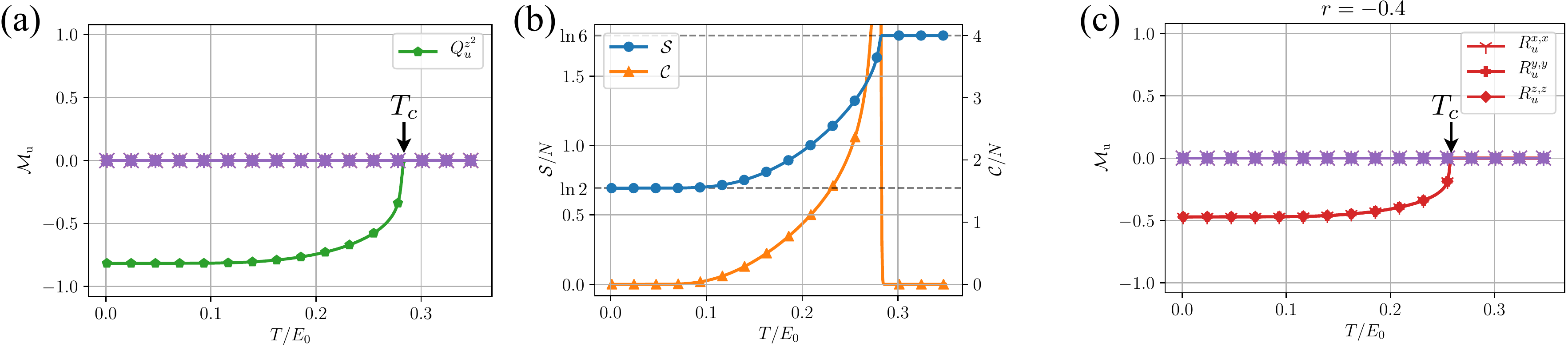}
    \caption{
        Temperature dependence of (a) the order parameter, (b, left axis) the entropy and (b, right axis) the specific heat for $n = 3$ uniform spherical model with $J/U=-0.1$.
        (c) Similar order-parameter plots for the $n = 3$ single-sublattice model with coupling constant ratio $r = -0.4$.
        The energy unit is $E_0 = t^2/U$.
    }
	\label{fig:temperature_n=3_spherical_J=-0.1_unif}
	\end{figure*}

Having the geometrically frustrated lattice in mind, we assume that the spatially modulated solutions are not realized.
Then we seek for the spatially uniform solutions (single-sublattice) only.

    Figure~\ref{fig:temperature_n=3_spherical_J=-0.1_unif}(a) shows the order parameter for the single-sublattice model with $n = 3$, $J/U = -0.1$.
    The system shows the $Q^{z^2}$ order at $T_c/E_0 \simeq 0.28$, which is consistent with the magnitude of the coupling constant shown in Fig.~\ref{fig:coupling_constants_n=3}.
    The entropy and specific heat are shown in Fig.~\ref{fig:temperature_n=3_spherical_J=-0.1_unif}(b) with left and right axis, respectively.
    The residual entropy $\mathcal{S} = \ln 2$ remains, which is in accordance with the degeneracy of spin in the absence of the sublattice degrees of freedom.
    Namely, the wave function of the ground state is degenerated and is written as
	\begin{align}
		\Ket{\vec{\psi}}
		=
		\mqty(
			\Ket{z, \ua} \\
			\Ket{z, \da}
		).
	\end{align}
	We have confirmed that the eigenvalues of $\hat a $ in Eq.~\eqref{eq:hessian} are all non-negative (not shown) and thus the ordered state is stable.

    We also point out the other interesting possibilities.
    The above orbital order is induced by the coupling constant $I_Q >0$ in Fig.~\ref{fig:coupling_constants_n=3}.
    In this figure, it is notable that the values of $I_Q$ and $I_R$ are very close with each other. 
    Then we try to search for another solutions by
    introducing the modified coupling constants defined as
    \begin{align}
        &\tilde{I}_Q
        =
        \qty(1 + r) I_Q,
        \\
        &\tilde{I}_R
        =
        \qty(1 - r) I_R,
    \end{align}
	where the original spherical model corresponds to $r = 0$.
	
	We show the order parameters for $n = 3$, $J/U = -0.1$ uniform model with the coupling constant ratio $r = -0.4$ in Fig.~\ref{fig:temperature_n=3_spherical_J=-0.1_unif}(c).
	Since the magnitude of the modified coupling constants satisfies $\tilde{I}_R > \tilde{I}_Q$ in the present condition, we obtain the solution for $R^{\mu,\mu}$ moments.
	Recalling the definition of the $R$ moment, we may rewrite the order parameter as $R^{\mu,\mu} \sim L^{\mu} S^{\mu}$ symbolically.
	Therefore, it is interpreted that the system has the effective spin-orbit coupling spontaneously.
	The wave function is written as
	\begin{align}
		\Ket{\vec{\psi}}
		=
		\frac{1}{\sqrt{3}} \mqty(
			\Ket{x,\ua} - \imu \Ket{y,\ua} - \Ket{z,\da} \\
			-\Ket{x,\da} - \imu \Ket{y,\da} - \Ket{z,\ua}
		),
	\end{align}
	which indicates that the ground state is entangled with respect to spin and orbital.
    These doubly degenerate ground states are connected with each other by the time-reversal symmetry.

This ``spontaneous spin-orbit coupling'' splits the six-fold degeneracy into two-fold and four-fold multiplets, and which is realized in the ground state is dependent on the sign of the order parameters.
Our solutions show that the ground state is always doubly degenerate, and this should be related to the minimization of the entropy at low temperatures.

Thus, although the system at the original parameter shows the doublon-orbital ordering ($Q$), the system is located near the parameter range where the intriguing $R$ order occurs.
As discussed in Sec.~\ref{sec:hubbard} the original spin-orbit coupling $\Lambda_{\rm SO}$ is tiny, but it might enter through the $R$-type ordering.
Such situation is realized only for $n=3$ model with the antiferromagnetic Hund's coupling.

\section{Numerical results for fullerides} \label{sec:results_fullerides}

We show the numerical results for the fulleride in the strong coupling regime by using the hopping parameters obtained by the first principles calculation~\cite{Nomura2012}.
We take the intra-orbital Coulomb interaction $U = 1\mathrm{eV}$ and the Hund's coupling $J/U=-0.1$ in the following.

\subsection{A15 structure} \label{sec:results_a15}

	\begin{figure}[t]
		\centering
		\includegraphics[width=85mm]{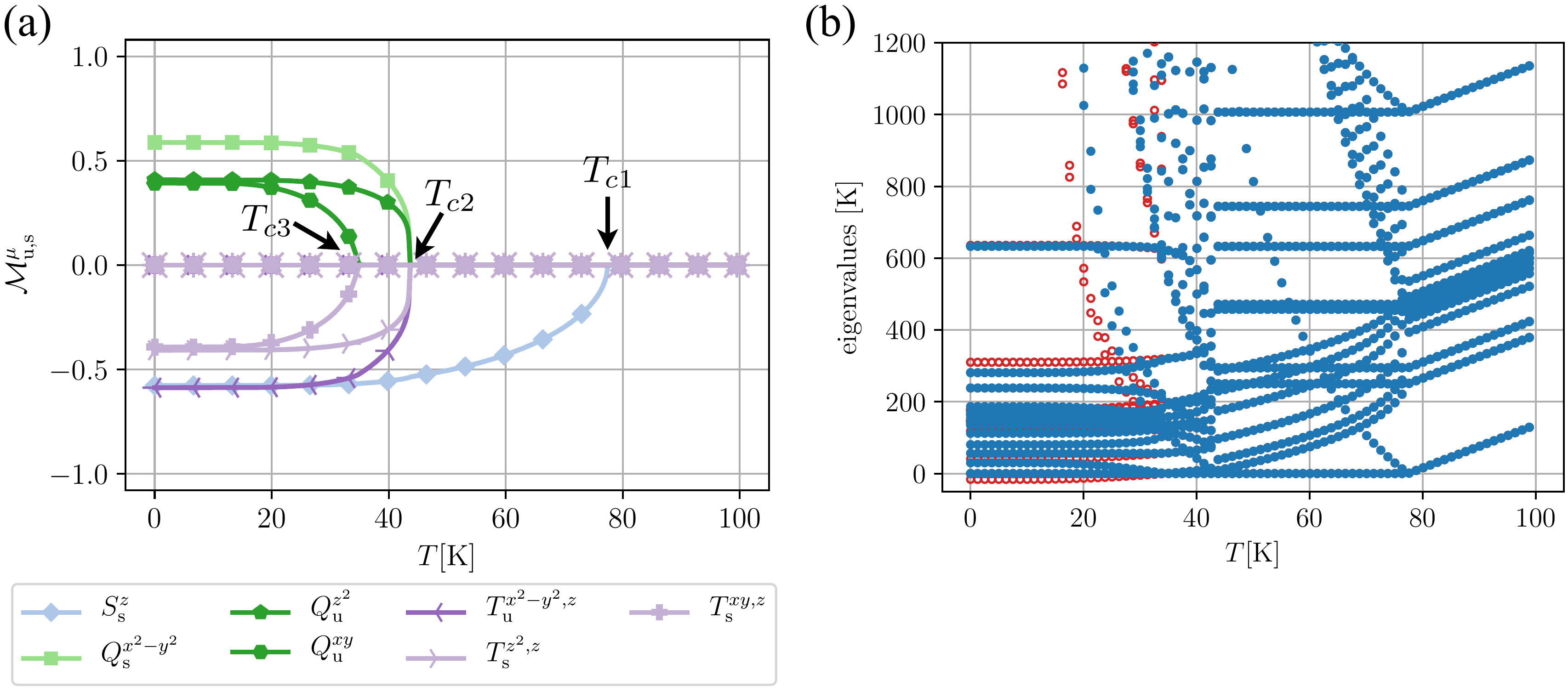}
		\caption{
			Temperature dependence of (a) the order parameter and (b) the eigenvalues of the matrix $\hat{a}$ for A15 fulleride model.
			The blue filled symbol in (b) corresponds to the solution given in (a).
			The red circle represents the solution without the phase transition at $T_{c3}$.
		}
		\label{fig:temperature_a15}
	\end{figure}

First of all we show in Fig.~\ref{fig:temperature_a15}(a) the temperature dependence of order parameters for the strong-coupling limit model of the realistic fulleride material with the A15 structure. The hopping parameters for $\mathrm{Cs}_3\mathrm{C}_{60}$ is chosen (A15-Cs($V_{\mathrm{SC}}^{\mathrm{opt-}P}$) in Ref.~\cite{Nomura2012}).
The lattice structure is a bipartite lattice, and A and B sublattices are connected with each other by screw transformation (i.e., translation plus four-fold rotation).
As shown in the figure, at $T_{c1}\simeq 80$K, the antiferromagnetic moment (AF-$S$) appears by the second-order phase transition.
At lower temperatures, we identify the two successive phase transitions ($T_{c2,3}$) with orbital moment $Q$ and spin-orbital moment $T$.
These two $Q,T$ moments share the same symmetry under the presence of AF-$S^z$ order.
We cannot simply conclude which one is the primary order parameter, because the interaction has complicated form for the realistic model and cannot be decomposed to each contribution as in the spherical model.
We also note that our choice of parameter is not fine-tuned to reproduce correctly the transition temperature in the actual materials, although our results can be compared with the experiments semi-quantitatively.

We show in Fig.~\ref{fig:temperature_a15}(b) the eigenvalues (filled blue symbols) of the Hessian matrix defined in Eq.~\eqref{eq:hessian}.
All the values are non-negative, and therefore the system is stable.
On the other hand, we can also calculate the low-temperature solutions by suppressing the ordering at $T_{c3}$.
The results are plotted as the open red symbols in Fig.~\ref{fig:temperature_a15}(b).
In this case, the eigenvalues become partially negative and hence the system is not stable although the entropy goes to zero even in this case.
Thus, the emergence of the order at $T_{c3}$ is essential in order to reach the stable ground state.

	\begin{figure}[t]
		\centering
		\includegraphics[width=85mm]{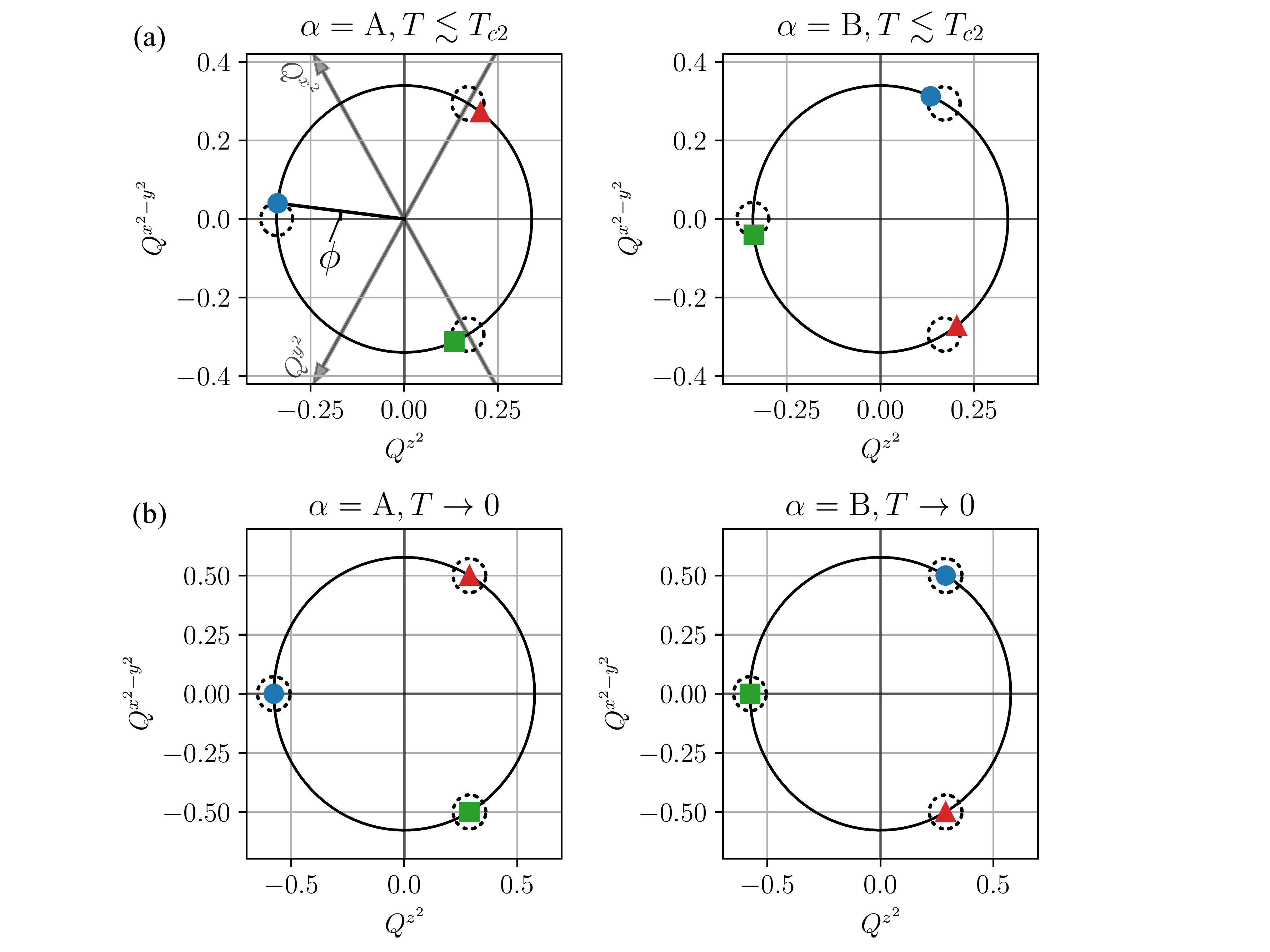}
		\caption{
			(a) Sublattice-dependent order parameters in the plane of $Q^{z^2}$-$Q^{x^2-y^2}$ for A (left) an B (right) sublattices at $T=40.4$K ($<T_{c2}$).
			The similar plots at low-temperature limit {\it without} the transition at $T_{c3}$ is shown in (b).
			The dashed circles in (a,b) correspond to the solutions in the system with four-fold symmetries.
Each color shows different kind of solutions, which share the same free energy.			
			The gray arrows with $Q^{x^2}$ or $Q^{y^2}$ are the guide for taking the other quantization axis.
			Specifically, the solution given in Fig.~\ref{fig:temperature_a15}(a) corresponds to the blue circle in the present figure (a).
			The angle $\phi$ in the left panel of (a) is the deviation from the horizontal axis.
		}
		\label{fig:ord_a15}
	\end{figure}

We discuss the origin of the second orbital order at $T_{c3}$ in more detail.
Below, we concentrate on the properties of $Q$ moments to make the discussion simple, since the symmetry of $Q$ is same as that of $T$ below the transition temperature $T_{c1}$.
Figure~\ref{fig:ord_a15}(a) shows the orbital order parameters for sublattice A (left panel) and B (right panel) slightly below the transition temperature $T_{c2}$ (but above $T_{c3}$).
The three patterns are obtained depending on the initial condition and hence are degenerate solutions.
It is seen from Fig.~\ref{fig:ord_a15}(a) that the plane of $X_\al=Q_\al^{z^2}$ and $Y_\al=Q_\al^{x^2-y^2}$ has a three-fold rotational symmetry and the equilateral triangle points, where the free energy minima are located, are tilted from the $X$ axis.
This tilt angle remains finite at low temperatures below $T_{c3}$.

This result can be understood from the Landau theory:
we can show that, without four-fold rotational symmetry as in $T_h$ point group symmetry in fulleride materials, the Landau free energy  is written in the restricted order-parameter space as
\begin{align}
\mathcal{F}_{\rm L} &= \sum_{\al={\rm A,B}} \Big[
c_1 X_\al(X_\al^2-3Y_\al^2) + c_2 s_\al Y_\al (3X_\al^2 -Y_\al^2)
\Big],
\end{align}
where $s_{\al =\rm A} = +1$ and $s_{\al =\rm B} = -1$.
We have considered only the third-order term for our purpose.
This is consistent with the numerical results and the tilt of the angle is due to the presence of $c_2$ term.
The tilt angle is estimated with the polar coordinates $X=r\cos \theta$ and $Y=r\sin \theta$, leading to another expression of the free energy
$\mathcal{F}_{\rm L} \propto \cos (3\theta + \phi)$ with $\phi =\tan^{-1} c_2/c_1$ being the tilt angle.
For example, one can estimate this angle from Fig.~\ref{fig:ord_a15}(a) as $\phi=6.76^\circ$.
The A15 structure has the screw symmetry, i.e., the combination of the translation along [111] and four-fold rotation around $x,y,z$ axes, which relates the order parameters at A and B sublattices.
Indeed, the above Landau free energy is invariant under the three-fold rotation and screw transformations.

If the four-fold symmetry is present, the condition $c_2 = 0$ or $\phi=0$ is required.
In Fig.~\ref{fig:ord_a15}(b), we show the order parameters at $T\to 0$ {\it without} the second orbital ordering below $T_{c3}$, where the four-fold symmetry seems to be effectively recovered since the tilt angle goes to zero when $T\to 0$.
Hence, the origin of the second orbital order in Fig.~\ref{fig:temperature_a15}(a) below $T_{c3}$ is interpreted as induced from this emergent symmetry at low temperatures which provides an additional free energy gain.

\subsection{fcc structure} \label{sec:results_fcc}

	\begin{figure}[t]
		\centering
		\includegraphics[width=85mm]{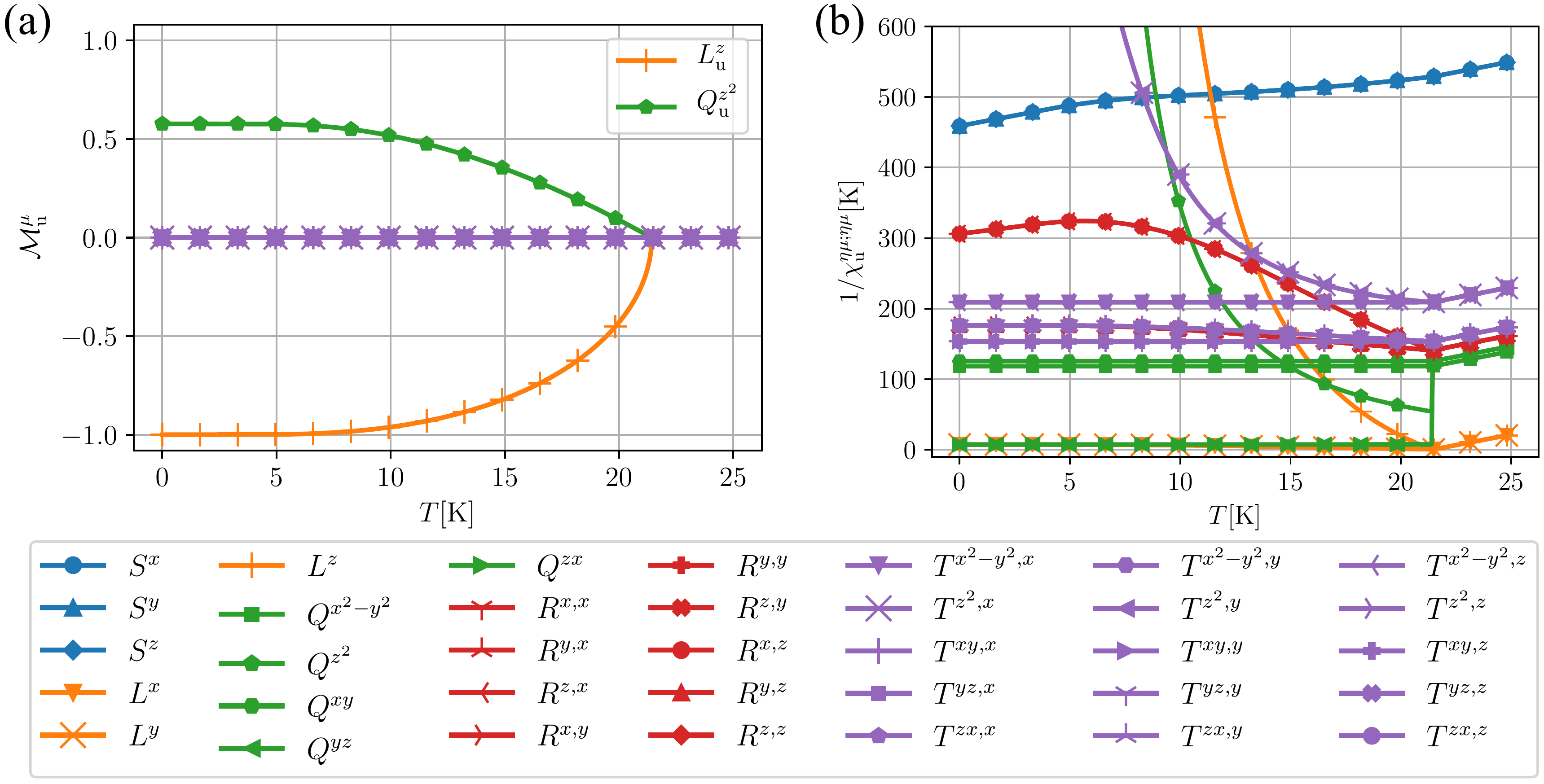}
		\caption{
			Temperature dependence of (a) the order parameter and (b) the inverse of the diagonal susceptibility for uniform fcc fulleride model with $J/U=-0.1$.
		}
	\label{fig:temperature_fcc}
	\end{figure}
	
Finally, we consider the fulleride material with the fcc structure.
The spin-orbital model in the strong-coupling limit is obtained by using the hopping parameters for $\mathrm{Rb}_3\mathrm{C}_{60}$ in Ref.~\cite{Nomura2012}.
Because of the geometrically frustrated nature of the fcc lattice, we here seek for only the spatially uniform ordered states.

Figure~\ref{fig:temperature_fcc}(a) shows the temperature evolution of the order parameters.
Here the primary order parameter is the uniform $L^z$ moment which breaks the time-reversal symmetry.
The $L^z$-order arises as $(T_c-T)^{1/2}$, and the $Q^{z^2}$ is also induced simultaneously with the linear temperature dependence $\propto (T_c-T)$.
The latter $Q$ moment is induced from the coupling term with the form $(L^z)^2Q^{z^2}$ in the Landau free energy.
We note that $L_z$ is not induced when $Q^{z^2}$ is a primary order parameter from that coupling, since $L_z$ and $Q^{z^2}$ have different time-reversal symmetry [$(L^z)^2$ and $Q^{z^2}$ are same].
The ground-state wave function is written in a simple form as
	\begin{align}
		\Ket{\vec{\psi}}
		=
		\frac{1}{\sqrt{2}} \mqty(
			\Ket{x,\ua} - \imu \Ket{y,\ua} \\
			\Ket{x,\da} - \imu \Ket{y,\da}
		),
	\end{align}
where the complex wave function clearly shows the time-reversal symmetry breaking.
We note that this orbital moment is not a simple orbital motion around the fullerene molecule, but a complex motion of the three electron state given in Eq.~\eqref{eq:ground_state_single_site}.
In our calculations the spin $S$ order does not occur and the ground state is doubly degenerate at each cite.
The stability of the solution is checked by the non-negative eigenvalues of the Hessian matrix.

The results found here is different from the spherical case discussed in Sec.~\ref{sec:results_n=3_unif}.
The difference is due to the specific form of the tight-binding hopping parameters.
We show one of the coupling constant $I_L$ for the nearest neighbour sites as
\begin{align}
    &\mqty(
        I_1 & I_2 & 0 \\
        I_2 & -I_3 & 0 \\
        0 & 0 & I_4 
    ), \ \ \bm{R} = \qty(\frac{a}{2}, \frac{a}{2}, 0),
\end{align}
where $\bm{R}$ is the direction of the NN molecules and $a$ is the lattice constant for the fcc fulleride.
The information for the other NN pairs is constructed from the symmetry operations.
The values of the matrix element are $I_1 = 12.5, I_2 = 9.77, I_3 = 0.511$ and $I_4 = 20.1$ in units of K in the present models.
The coupling constant has the same symmetry as the hopping parameters in Ref.~\cite{Nomura2012} as required by the space group symmetry.
The nearest neighbor coupling constant is largest and is positive, which favors the uniform magnetic orbital moment $\bm L$.
As for the next nearest neighbour site, the coupling constant matrices are diagonal and every component of them is smaller than nearest neighbour ones.

Since the spin $S$ moment has the same symmetry as $L$, it can in general be simultaneously induced under the small but finite spin-orbit coupling.
However, as discussed in Sec.~\ref{sec:hubbard}, the magnitude of the effective spin-orbit coupling for the doublon orbital is $\Lambda_{\rm SO}\sim 10^{-9}$eV, which can be regarded as zero in practice.
Hence, the spin order can occur independently at low temperatures.
The absence of the spin $S$ order is interpreted from the point of view of the coupling constant.
Figure~\ref{fig:temperature_fcc}(b) shows that the temperature dependence of the inverse of the diagonal susceptibilities.
The blue lines represents the magnetic susceptibility ($S$), which indicates that the coupling constants of $S$ are antiferromagnetic owing to the negative Curie-Weiss temperature.
In this case, the transition temperature should be very low due to the geometrical frustration of fcc lattice, but finally the system should show some magnetic ordering~\cite{Kasahara2014}.

\subsection{Discussion}

The models in this section are based on the band-structure calculation results.
Furthermore, the fulleride materials can be located in the Mott insulator regime depending on the pressure.
Hence, our results are potentially applied to the real materials.
In fulleride materials, the antiferromagnetic orders is experimentally identified at low temperatures, while the orbital orders are not yet reported.
Based on our results, we propose that at low temperatures the orbital ordered moments $Q$ are induced with two successive transitions for A15 structures, and $L$ moments may appear for fcc structures.
Such fingerprints of the orbital orders may be found in thermodynamic quantities in principle.
Here the orbital moment is not for a usual electron but for the doublons specific to the systems with antiferromagnetic Hund's coupling as emphasized in the present paper.
On the other hand, since the real compounds are polycrystals and the disorder effects are also present, the orbital orders might be smeared out in realistic situations.
In this context, the effect of disorders on our spin-orbital model is interesting future issues which make it more direct to compare the theoretical results with experimental observations.
Moreover, the antiferromagnetic Hund's coupling originates from the electron-phonon coupling.
The resultant retardation effects are also the parts not included in this paper and an important issue for the more realistic arguments.

\section{Summary and Outlook} \label{sec:summary}
In order to clarify the properties of strongly correlated electrons in fulleride superconductors, we have constructed the spin-orbital model in the strong coupling limit.
We begin with the three-orbital Hubbard model with the antiferromagnetic Hund's coupling which is realized by the coupling between the electronic degrees of freedom and anisotropic Jahn-Teller molecular vibrations.
In this case, the pair hopping effect among the different orbitals becomes relevant in strong contrast to the multiorbital $d$-electron systems with the ferromagnetic Hund's coupling.
We have mainly considered the half-filled $n=3$ case relevant to real materials, where it is composed of the singly-occupied (singlon) plus doubly occupied orbitals (doublon) as illustrated in Fig.~\ref{fig:ground_state}.
The correlated ground state for an isolated fullerene molecule is six-fold degenerate and is characterized by the spin and orbital indices.
This is the situation similar to the $n=1$ ground states and the analogy between $n=3$ and $n=1$ helps us for interpreting the results.
The usual orbital moment, which is present for the $n=1$ case, is absent for $n=3$ because of the correlated nature of the wave function, and instead the active orbital moment characteristic for doublons exists. 
As the result, the spin-orbit coupling, which is the order of 1meV for $p$-electrons, becomes 1neV because of the extended nature of the molecular orbitals and the correlation effects.

We have applied the second-order perturbation theory with respect to the inter-molecule hopping, and have obtained the localized spin-orbital model specific to the fullerides.
The obtained spin-orbital model is analyzed by employing the mean-field approximation.
For reference, we have first solved the spherical $n=1$ model for both ferromagnetic and antiferromagnetic Hund's couplings with a spherical limit for the bipartite lattice.
We then apply our method to the $n=3$ model where the magnetic order is found at relatively high temperatures and the orbital order also occurs at lower temperatures.
The temperature dependences of the physical quantities such as order parameters,  internal and free energies, specific heat, entropy, and susceptibilities are investigated in detail.
The thermodynamic stability is also studied based on the Hessian matrix derived from the inverse susceptibilities, and are checked by confirming that all the eigenvalues are non-negative.

We have also considered the realistic situation in alkali-doped fullerides, by using the tight-binding parameters derived from the first principles calculations.
For the choice of the lattice structure, we have taken both the bipartite A15 and fcc structures, whose hopping parameters have been derived in Ref.~\cite{Nomura2012}.
For the A15 structure, the antiferromagnetic order occurs at high temperatures, and the electric orbital orders arise at lower temperatures with two successive transitions.
The first orbital order is already captured in the spherical model, but the second orbital order is characteristic for the $T_h$ symmetry in fulleride materials where only the three-fold rotation symmetry exists.
This point has been discussed in detail based on the Landau theory.
For the fcc model, we have concentrated on the spatially uniform solutions due to the geometrically frustrated nature of the lattice. 
We have found that the magnetic orbital order occurs.
Although this orbital moment has the same symmetry as the electronic spin, the spin moment is not induced simultaneously in fulleride since the spin-orbit coupling is tiny as mentioned above.
Thus the spin-moment can order independently, and is expected to be antiferromagnetically ordered in the ground state where the transition temperature is expected to be low owing to the geometrical frustration of the fcc lattice.

Our formalism itself is constructed in a very general way, and can be applied to any systems in the strong coupling limit with integer fillings per atom or molecule.
In this context, it would be desirable to develop the general framework for the strong-coupling-limit spin-orbital model with the combination of the hopping parameters in the Wannier functions obtained from the band-structure calculations.
This application is of interest specifically in studying the ordered state of the multiorbital electronic systems including transition metals and organic materials.
This point remains to be explored and is an intriguing issue in the future.

\section*{Acknowledgement}
This work was supported by JSPS KAKENHI Grants
No. JP18K13490
and
No. JP19H01842.
\bibliographystyle{apsrev4-2}
\bibliography{fulleride.bbl}

\begin{thebibliography}{41}%
\makeatletter
\providecommand \@ifxundefined [1]{%
 \@ifx{#1\undefined}
}%
\providecommand \@ifnum [1]{%
 \ifnum #1\expandafter \@firstoftwo
 \else \expandafter \@secondoftwo
 \fi
}%
\providecommand \@ifx [1]{%
 \ifx #1\expandafter \@firstoftwo
 \else \expandafter \@secondoftwo
 \fi
}%
\providecommand \natexlab [1]{#1}%
\providecommand \enquote  [1]{``#1''}%
\providecommand \bibnamefont  [1]{#1}%
\providecommand \bibfnamefont [1]{#1}%
\providecommand \citenamefont [1]{#1}%
\providecommand \href@noop [0]{\@secondoftwo}%
\providecommand \href [0]{\begingroup \@sanitize@url \@href}%
\providecommand \@href[1]{\@@startlink{#1}\@@href}%
\providecommand \@@href[1]{\endgroup#1\@@endlink}%
\providecommand \@sanitize@url [0]{\catcode `\\12\catcode `\$12\catcode
  `\&12\catcode `\#12\catcode `\^12\catcode `\_12\catcode `\%12\relax}%
\providecommand \@@startlink[1]{}%
\providecommand \@@endlink[0]{}%
\providecommand \url  [0]{\begingroup\@sanitize@url \@url }%
\providecommand \@url [1]{\endgroup\@href {#1}{\urlprefix }}%
\providecommand \urlprefix  [0]{URL }%
\providecommand \Eprint [0]{\href }%
\providecommand \doibase [0]{https://doi.org/}%
\providecommand \selectlanguage [0]{\@gobble}%
\providecommand \bibinfo  [0]{\@secondoftwo}%
\providecommand \bibfield  [0]{\@secondoftwo}%
\providecommand \translation [1]{[#1]}%
\providecommand \BibitemOpen [0]{}%
\providecommand \bibitemStop [0]{}%
\providecommand \bibitemNoStop [0]{.\EOS\space}%
\providecommand \EOS [0]{\spacefactor3000\relax}%
\providecommand \BibitemShut  [1]{\csname bibitem#1\endcsname}%
\let\auto@bib@innerbib\@empty
%</preamble>
\bibitem [{\citenamefont {Hebard}\ \emph {et~al.}(1991)\citenamefont {Hebard},
  \citenamefont {Rosseinsky}, \citenamefont {Haddon}, \citenamefont {Murphy},
  \citenamefont {Glarum}, \citenamefont {Palstra}, \citenamefont {Ramirez},\
  and\ \citenamefont {Kortan}}]{Hebard1991}%
  \BibitemOpen
  \bibfield  {author} {\bibinfo {author} {\bibfnamefont {A.~F.}\ \bibnamefont
  {Hebard}}, \bibinfo {author} {\bibfnamefont {M.~J.}\ \bibnamefont
  {Rosseinsky}}, \bibinfo {author} {\bibfnamefont {R.~C.}\ \bibnamefont
  {Haddon}}, \bibinfo {author} {\bibfnamefont {D.~W.}\ \bibnamefont {Murphy}},
  \bibinfo {author} {\bibfnamefont {S.~H.}\ \bibnamefont {Glarum}}, \bibinfo
  {author} {\bibfnamefont {T.~T.~M.}\ \bibnamefont {Palstra}}, \bibinfo
  {author} {\bibfnamefont {A.~P.}\ \bibnamefont {Ramirez}},\ and\ \bibinfo
  {author} {\bibfnamefont {A.~R.}\ \bibnamefont {Kortan}},\ }\href
  {https://doi.org/10.1038/350600a0} {\bibfield  {journal} {\bibinfo  {journal}
  {Nature (London)}\ }\textbf {\bibinfo {volume} {350}},\ \bibinfo {pages}
  {600} (\bibinfo {year} {1991})}\BibitemShut {NoStop}%
\bibitem [{\citenamefont {Rosseinsky}\ \emph {et~al.}(1991)\citenamefont
  {Rosseinsky}, \citenamefont {Ramirez}, \citenamefont {Glarum}, \citenamefont
  {Murphy}, \citenamefont {Haddon}, \citenamefont {Hebard}, \citenamefont
  {Palstra}, \citenamefont {Kortan}, \citenamefont {Zahurak},\ and\
  \citenamefont {Makhija}}]{Rosseinsky1991}%
  \BibitemOpen
  \bibfield  {author} {\bibinfo {author} {\bibfnamefont {M.~J.}\ \bibnamefont
  {Rosseinsky}}, \bibinfo {author} {\bibfnamefont {A.~P.}\ \bibnamefont
  {Ramirez}}, \bibinfo {author} {\bibfnamefont {S.~H.}\ \bibnamefont {Glarum}},
  \bibinfo {author} {\bibfnamefont {D.~W.}\ \bibnamefont {Murphy}}, \bibinfo
  {author} {\bibfnamefont {R.~C.}\ \bibnamefont {Haddon}}, \bibinfo {author}
  {\bibfnamefont {A.~F.}\ \bibnamefont {Hebard}}, \bibinfo {author}
  {\bibfnamefont {T.~T.~M.}\ \bibnamefont {Palstra}}, \bibinfo {author}
  {\bibfnamefont {A.~R.}\ \bibnamefont {Kortan}}, \bibinfo {author}
  {\bibfnamefont {S.~M.}\ \bibnamefont {Zahurak}},\ and\ \bibinfo {author}
  {\bibfnamefont {A.~V.}\ \bibnamefont {Makhija}},\ }\href
  {https://doi.org/10.1103/PhysRevLett.66.2830} {\bibfield  {journal} {\bibinfo
   {journal} {Phys. Rev. Lett.}\ }\textbf {\bibinfo {volume} {66}},\ \bibinfo
  {pages} {2830} (\bibinfo {year} {1991})}\BibitemShut {NoStop}%
\bibitem [{\citenamefont {Holczer}\ \emph {et~al.}(1991)\citenamefont
  {Holczer}, \citenamefont {Klein}, \citenamefont {mei Huang}, \citenamefont
  {Kaner}, \citenamefont {jian Fu}, \citenamefont {Whetten},\ and\
  \citenamefont {Diederich}}]{Holczer1991}%
  \BibitemOpen
  \bibfield  {author} {\bibinfo {author} {\bibfnamefont {K.}~\bibnamefont
  {Holczer}}, \bibinfo {author} {\bibfnamefont {O.}~\bibnamefont {Klein}},
  \bibinfo {author} {\bibfnamefont {S.}~\bibnamefont {mei Huang}}, \bibinfo
  {author} {\bibfnamefont {R.~B.}\ \bibnamefont {Kaner}}, \bibinfo {author}
  {\bibfnamefont {K.}~\bibnamefont {jian Fu}}, \bibinfo {author} {\bibfnamefont
  {R.~L.}\ \bibnamefont {Whetten}},\ and\ \bibinfo {author} {\bibfnamefont
  {F.}~\bibnamefont {Diederich}},\ }\href
  {https://doi.org/10.1126/science.252.5009.1154} {\bibfield  {journal}
  {\bibinfo  {journal} {Science}\ }\textbf {\bibinfo {volume} {252}},\ \bibinfo
  {pages} {1154} (\bibinfo {year} {1991})}\BibitemShut {NoStop}%
\bibitem [{\citenamefont {Tanigaki}\ \emph {et~al.}(1991)\citenamefont
  {Tanigaki}, \citenamefont {Ebbesen}, \citenamefont {Saito}, \citenamefont
  {Mizuki}, \citenamefont {Tsai}, \citenamefont {Kubo},\ and\ \citenamefont
  {Kuroshima}}]{Tanigaki1991}%
  \BibitemOpen
  \bibfield  {author} {\bibinfo {author} {\bibfnamefont {K.}~\bibnamefont
  {Tanigaki}}, \bibinfo {author} {\bibfnamefont {T.~W.}\ \bibnamefont
  {Ebbesen}}, \bibinfo {author} {\bibfnamefont {S.}~\bibnamefont {Saito}},
  \bibinfo {author} {\bibfnamefont {J.}~\bibnamefont {Mizuki}}, \bibinfo
  {author} {\bibfnamefont {J.~S.}\ \bibnamefont {Tsai}}, \bibinfo {author}
  {\bibfnamefont {Y.}~\bibnamefont {Kubo}},\ and\ \bibinfo {author}
  {\bibfnamefont {S.}~\bibnamefont {Kuroshima}},\ }\href
  {https://doi.org/10.1038/352222a0} {\bibfield  {journal} {\bibinfo  {journal}
  {Nature (London)}\ }\textbf {\bibinfo {volume} {352}},\ \bibinfo {pages}
  {222} (\bibinfo {year} {1991})}\BibitemShut {NoStop}%
\bibitem [{\citenamefont {Fleming}\ \emph {et~al.}(1991)\citenamefont
  {Fleming}, \citenamefont {Ramirez}, \citenamefont {Rosseinsky}, \citenamefont
  {Murphy}, \citenamefont {Haddon}, \citenamefont {Zahurak},\ and\
  \citenamefont {Makhija}}]{Fleming1991}%
  \BibitemOpen
  \bibfield  {author} {\bibinfo {author} {\bibfnamefont {R.~M.}\ \bibnamefont
  {Fleming}}, \bibinfo {author} {\bibfnamefont {A.~P.}\ \bibnamefont
  {Ramirez}}, \bibinfo {author} {\bibfnamefont {M.~J.}\ \bibnamefont
  {Rosseinsky}}, \bibinfo {author} {\bibfnamefont {D.~W.}\ \bibnamefont
  {Murphy}}, \bibinfo {author} {\bibfnamefont {R.~C.}\ \bibnamefont {Haddon}},
  \bibinfo {author} {\bibfnamefont {S.~M.}\ \bibnamefont {Zahurak}},\ and\
  \bibinfo {author} {\bibfnamefont {A.~V.}\ \bibnamefont {Makhija}},\ }\href
  {https://doi.org/10.1038/352787a0} {\bibfield  {journal} {\bibinfo  {journal}
  {Nature (London)}\ }\textbf {\bibinfo {volume} {352}},\ \bibinfo {pages}
  {787} (\bibinfo {year} {1991})}\BibitemShut {NoStop}%
\bibitem [{\citenamefont {Ganin}\ \emph {et~al.}(2008)\citenamefont {Ganin},
  \citenamefont {Takabayashi}, \citenamefont {Khimyak}, \citenamefont
  {Margadonna}, \citenamefont {Tamai}, \citenamefont {Rosseinsky},\ and\
  \citenamefont {Prassides}}]{Ganin2008}%
  \BibitemOpen
  \bibfield  {author} {\bibinfo {author} {\bibfnamefont {A.~Y.}\ \bibnamefont
  {Ganin}}, \bibinfo {author} {\bibfnamefont {Y.}~\bibnamefont {Takabayashi}},
  \bibinfo {author} {\bibfnamefont {Y.~Z.}\ \bibnamefont {Khimyak}}, \bibinfo
  {author} {\bibfnamefont {S.}~\bibnamefont {Margadonna}}, \bibinfo {author}
  {\bibfnamefont {A.}~\bibnamefont {Tamai}}, \bibinfo {author} {\bibfnamefont
  {M.~J.}\ \bibnamefont {Rosseinsky}},\ and\ \bibinfo {author} {\bibfnamefont
  {K.}~\bibnamefont {Prassides}},\ }\href {https://doi.org/10.1038/nmat2179}
  {\bibfield  {journal} {\bibinfo  {journal} {Nat. Mater.}\ }\textbf {\bibinfo
  {volume} {7}},\ \bibinfo {pages} {367} (\bibinfo {year} {2008})}\BibitemShut
  {NoStop}%
\bibitem [{\citenamefont {Takabayashi}\ \emph {et~al.}(2009)\citenamefont
  {Takabayashi}, \citenamefont {Ganin}, \citenamefont {Jeglič}, \citenamefont
  {Arčon}, \citenamefont {Takano}, \citenamefont {Iwasa}, \citenamefont
  {Ohishi}, \citenamefont {Takata}, \citenamefont {Takeshita}, \citenamefont
  {Prassides},\ and\ \citenamefont {Rosseinsky}}]{Takabayashi2009}%
  \BibitemOpen
  \bibfield  {author} {\bibinfo {author} {\bibfnamefont {Y.}~\bibnamefont
  {Takabayashi}}, \bibinfo {author} {\bibfnamefont {A.~Y.}\ \bibnamefont
  {Ganin}}, \bibinfo {author} {\bibfnamefont {P.}~\bibnamefont {Jeglič}},
  \bibinfo {author} {\bibfnamefont {D.}~\bibnamefont {Arčon}}, \bibinfo
  {author} {\bibfnamefont {T.}~\bibnamefont {Takano}}, \bibinfo {author}
  {\bibfnamefont {Y.}~\bibnamefont {Iwasa}}, \bibinfo {author} {\bibfnamefont
  {Y.}~\bibnamefont {Ohishi}}, \bibinfo {author} {\bibfnamefont
  {M.}~\bibnamefont {Takata}}, \bibinfo {author} {\bibfnamefont
  {N.}~\bibnamefont {Takeshita}}, \bibinfo {author} {\bibfnamefont
  {K.}~\bibnamefont {Prassides}},\ and\ \bibinfo {author} {\bibfnamefont
  {M.~J.}\ \bibnamefont {Rosseinsky}},\ }\href
  {https://doi.org/10.1126/science.1169163} {\bibfield  {journal} {\bibinfo
  {journal} {Science}\ }\textbf {\bibinfo {volume} {323}},\ \bibinfo {pages}
  {1585} (\bibinfo {year} {2009})}\BibitemShut {NoStop}%
\bibitem [{\citenamefont {Ganin}\ \emph {et~al.}(2010)\citenamefont {Ganin},
  \citenamefont {Takabayashi}, \citenamefont {Jeglič}, \citenamefont {Arčon},
  \citenamefont {Potočnik}, \citenamefont {Baker}, \citenamefont {Ohishi},
  \citenamefont {McDonald}, \citenamefont {Tzirakis}, \citenamefont {McLennan},
  \citenamefont {Darling}, \citenamefont {Takata}, \citenamefont {Rosseinsky},\
  and\ \citenamefont {Prassides}}]{Ganin2010}%
  \BibitemOpen
  \bibfield  {author} {\bibinfo {author} {\bibfnamefont {A.~Y.}\ \bibnamefont
  {Ganin}}, \bibinfo {author} {\bibfnamefont {Y.}~\bibnamefont {Takabayashi}},
  \bibinfo {author} {\bibfnamefont {P.}~\bibnamefont {Jeglič}}, \bibinfo
  {author} {\bibfnamefont {D.}~\bibnamefont {Arčon}}, \bibinfo {author}
  {\bibfnamefont {A.}~\bibnamefont {Potočnik}}, \bibinfo {author}
  {\bibfnamefont {P.~J.}\ \bibnamefont {Baker}}, \bibinfo {author}
  {\bibfnamefont {Y.}~\bibnamefont {Ohishi}}, \bibinfo {author} {\bibfnamefont
  {M.~T.}\ \bibnamefont {McDonald}}, \bibinfo {author} {\bibfnamefont {M.~D.}\
  \bibnamefont {Tzirakis}}, \bibinfo {author} {\bibfnamefont {A.}~\bibnamefont
  {McLennan}}, \bibinfo {author} {\bibfnamefont {G.~R.}\ \bibnamefont
  {Darling}}, \bibinfo {author} {\bibfnamefont {M.}~\bibnamefont {Takata}},
  \bibinfo {author} {\bibfnamefont {M.~J.}\ \bibnamefont {Rosseinsky}},\ and\
  \bibinfo {author} {\bibfnamefont {K.}~\bibnamefont {Prassides}},\ }\href
  {https://doi.org/10.1038/nature09120} {\bibfield  {journal} {\bibinfo
  {journal} {Nature (London)}\ }\textbf {\bibinfo {volume} {466}},\ \bibinfo
  {pages} {221} (\bibinfo {year} {2010})}\BibitemShut {NoStop}%
\bibitem [{\citenamefont {Gunnarsson}(1997)}]{Gunnarsson1997}%
  \BibitemOpen
  \bibfield  {author} {\bibinfo {author} {\bibfnamefont {O.}~\bibnamefont
  {Gunnarsson}},\ }\href {https://doi.org/10.1103/RevModPhys.69.575} {\bibfield
   {journal} {\bibinfo  {journal} {Rev. Mod. Phys.}\ }\textbf {\bibinfo
  {volume} {69}},\ \bibinfo {pages} {575} (\bibinfo {year} {1997})}\BibitemShut
  {NoStop}%
\bibitem [{\citenamefont {Fabrizio}\ and\ \citenamefont
  {Tosatti}(1997)}]{Fabrizio1997}%
  \BibitemOpen
  \bibfield  {author} {\bibinfo {author} {\bibfnamefont {M.}~\bibnamefont
  {Fabrizio}}\ and\ \bibinfo {author} {\bibfnamefont {E.}~\bibnamefont
  {Tosatti}},\ }\href {https://doi.org/10.1103/PhysRevB.55.13465} {\bibfield
  {journal} {\bibinfo  {journal} {Phys. Rev. B}\ }\textbf {\bibinfo {volume}
  {55}},\ \bibinfo {pages} {13465} (\bibinfo {year} {1997})}\BibitemShut
  {NoStop}%
\bibitem [{\citenamefont {Capone}\ \emph {et~al.}(2002)\citenamefont {Capone},
  \citenamefont {Fabrizio}, \citenamefont {Castellani},\ and\ \citenamefont
  {Tosatti}}]{Capone2002}%
  \BibitemOpen
  \bibfield  {author} {\bibinfo {author} {\bibfnamefont {M.}~\bibnamefont
  {Capone}}, \bibinfo {author} {\bibfnamefont {M.}~\bibnamefont {Fabrizio}},
  \bibinfo {author} {\bibfnamefont {C.}~\bibnamefont {Castellani}},\ and\
  \bibinfo {author} {\bibfnamefont {E.}~\bibnamefont {Tosatti}},\ }\href
  {https://doi.org/10.1126/science.1071122} {\bibfield  {journal} {\bibinfo
  {journal} {Science}\ }\textbf {\bibinfo {volume} {296}},\ \bibinfo {pages}
  {2364} (\bibinfo {year} {2002})}\BibitemShut {NoStop}%
\bibitem [{\citenamefont {Capone}\ \emph {et~al.}(2009)\citenamefont {Capone},
  \citenamefont {Fabrizio}, \citenamefont {Castellani},\ and\ \citenamefont
  {Tosatti}}]{Capone2009}%
  \BibitemOpen
  \bibfield  {author} {\bibinfo {author} {\bibfnamefont {M.}~\bibnamefont
  {Capone}}, \bibinfo {author} {\bibfnamefont {M.}~\bibnamefont {Fabrizio}},
  \bibinfo {author} {\bibfnamefont {C.}~\bibnamefont {Castellani}},\ and\
  \bibinfo {author} {\bibfnamefont {E.}~\bibnamefont {Tosatti}},\ }\href
  {https://doi.org/10.1103/RevModPhys.81.943} {\bibfield  {journal} {\bibinfo
  {journal} {Rev. Mod. Phys.}\ }\textbf {\bibinfo {volume} {81}},\ \bibinfo
  {pages} {943} (\bibinfo {year} {2009})}\BibitemShut {NoStop}%
\bibitem [{\citenamefont {Nomura}\ \emph {et~al.}(2016)\citenamefont {Nomura},
  \citenamefont {Sakai}, \citenamefont {Capone},\ and\ \citenamefont
  {Arita}}]{Nomura2016}%
  \BibitemOpen
  \bibfield  {author} {\bibinfo {author} {\bibfnamefont {Y.}~\bibnamefont
  {Nomura}}, \bibinfo {author} {\bibfnamefont {S.}~\bibnamefont {Sakai}},
  \bibinfo {author} {\bibfnamefont {M.}~\bibnamefont {Capone}},\ and\ \bibinfo
  {author} {\bibfnamefont {R.}~\bibnamefont {Arita}},\ }\href
  {https://doi.org/10.1088/0953-8984/28/15/153001} {\bibfield  {journal}
  {\bibinfo  {journal} {J. Phys. Condens. Matter}\ }\textbf {\bibinfo {volume}
  {28}},\ \bibinfo {pages} {153001} (\bibinfo {year} {2016})}\BibitemShut
  {NoStop}%
\bibitem [{\citenamefont {Takabayashi}\ and\ \citenamefont
  {Prassides}(2016)}]{Takabayashi2016}%
  \BibitemOpen
  \bibfield  {author} {\bibinfo {author} {\bibfnamefont {Y.}~\bibnamefont
  {Takabayashi}}\ and\ \bibinfo {author} {\bibfnamefont {K.}~\bibnamefont
  {Prassides}},\ }\href {https://doi.org/10.1098/rsta.2015.0320} {\bibfield
  {journal} {\bibinfo  {journal} {Phil. Trans. R. Soc. A}\ }\textbf {\bibinfo
  {volume} {374}},\ \bibinfo {pages} {20150320} (\bibinfo {year}
  {2016})}\BibitemShut {NoStop}%
\bibitem [{\citenamefont {Zadik}\ \emph {et~al.}(2015)\citenamefont {Zadik},
  \citenamefont {Takabayashi}, \citenamefont {Klupp}, \citenamefont {Colman},
  \citenamefont {Ganin}, \citenamefont {Potočnik}, \citenamefont {Jeglič},
  \citenamefont {Arčon}, \citenamefont {Matus}, \citenamefont {Kamarás},
  \citenamefont {Kasahara}, \citenamefont {Iwasa}, \citenamefont {Fitch},
  \citenamefont {Ohishi}, \citenamefont {Garbarino}, \citenamefont {Kato},
  \citenamefont {Rosseinsky},\ and\ \citenamefont {Prassides}}]{Zadik2015}%
  \BibitemOpen
  \bibfield  {author} {\bibinfo {author} {\bibfnamefont {R.~H.}\ \bibnamefont
  {Zadik}}, \bibinfo {author} {\bibfnamefont {Y.}~\bibnamefont {Takabayashi}},
  \bibinfo {author} {\bibfnamefont {G.}~\bibnamefont {Klupp}}, \bibinfo
  {author} {\bibfnamefont {R.~H.}\ \bibnamefont {Colman}}, \bibinfo {author}
  {\bibfnamefont {A.~Y.}\ \bibnamefont {Ganin}}, \bibinfo {author}
  {\bibfnamefont {A.}~\bibnamefont {Potočnik}}, \bibinfo {author}
  {\bibfnamefont {P.}~\bibnamefont {Jeglič}}, \bibinfo {author} {\bibfnamefont
  {D.}~\bibnamefont {Arčon}}, \bibinfo {author} {\bibfnamefont
  {P.}~\bibnamefont {Matus}}, \bibinfo {author} {\bibfnamefont
  {K.}~\bibnamefont {Kamarás}}, \bibinfo {author} {\bibfnamefont
  {Y.}~\bibnamefont {Kasahara}}, \bibinfo {author} {\bibfnamefont
  {Y.}~\bibnamefont {Iwasa}}, \bibinfo {author} {\bibfnamefont {A.~N.}\
  \bibnamefont {Fitch}}, \bibinfo {author} {\bibfnamefont {Y.}~\bibnamefont
  {Ohishi}}, \bibinfo {author} {\bibfnamefont {G.}~\bibnamefont {Garbarino}},
  \bibinfo {author} {\bibfnamefont {K.}~\bibnamefont {Kato}}, \bibinfo {author}
  {\bibfnamefont {M.~J.}\ \bibnamefont {Rosseinsky}},\ and\ \bibinfo {author}
  {\bibfnamefont {K.}~\bibnamefont {Prassides}},\ }\href
  {https://doi.org/10.1126/sciadv.1500059} {\bibfield  {journal} {\bibinfo
  {journal} {Sci. Adv.}\ }\textbf {\bibinfo {volume} {1}},\ \bibinfo {pages}
  {e1500059} (\bibinfo {year} {2015})}\BibitemShut {NoStop}%
\bibitem [{\citenamefont {Kasahara}\ \emph {et~al.}(2017)\citenamefont
  {Kasahara}, \citenamefont {Takeuchi}, \citenamefont {Zadik}, \citenamefont
  {Takabayashi}, \citenamefont {Colman}, \citenamefont {McDonald},
  \citenamefont {Rosseinsky}, \citenamefont {Prassides},\ and\ \citenamefont
  {Iwasa}}]{Kasahara2017}%
  \BibitemOpen
  \bibfield  {author} {\bibinfo {author} {\bibfnamefont {Y.}~\bibnamefont
  {Kasahara}}, \bibinfo {author} {\bibfnamefont {Y.}~\bibnamefont {Takeuchi}},
  \bibinfo {author} {\bibfnamefont {R.~H.}\ \bibnamefont {Zadik}}, \bibinfo
  {author} {\bibfnamefont {Y.}~\bibnamefont {Takabayashi}}, \bibinfo {author}
  {\bibfnamefont {R.~H.}\ \bibnamefont {Colman}}, \bibinfo {author}
  {\bibfnamefont {R.~D.}\ \bibnamefont {McDonald}}, \bibinfo {author}
  {\bibfnamefont {M.~J.}\ \bibnamefont {Rosseinsky}}, \bibinfo {author}
  {\bibfnamefont {K.}~\bibnamefont {Prassides}},\ and\ \bibinfo {author}
  {\bibfnamefont {Y.}~\bibnamefont {Iwasa}},\ }\href
  {https://doi.org/10.1038/ncomms14467} {\bibfield  {journal} {\bibinfo
  {journal} {Nat. Commun.}\ }\textbf {\bibinfo {volume} {8}},\ \bibinfo {pages}
  {14467} (\bibinfo {year} {2017})}\BibitemShut {NoStop}%
\bibitem [{\citenamefont {Han}\ \emph {et~al.}(2020)\citenamefont {Han},
  \citenamefont {Guan}, \citenamefont {Song}, \citenamefont {Wang},
  \citenamefont {Ren}, \citenamefont {Meng}, \citenamefont {Ma},\ and\
  \citenamefont {Xue}}]{Han2020}%
  \BibitemOpen
  \bibfield  {author} {\bibinfo {author} {\bibfnamefont {S.}~\bibnamefont
  {Han}}, \bibinfo {author} {\bibfnamefont {M.-X.}\ \bibnamefont {Guan}},
  \bibinfo {author} {\bibfnamefont {C.-L.}\ \bibnamefont {Song}}, \bibinfo
  {author} {\bibfnamefont {Y.-L.}\ \bibnamefont {Wang}}, \bibinfo {author}
  {\bibfnamefont {M.-Q.}\ \bibnamefont {Ren}}, \bibinfo {author} {\bibfnamefont
  {S.}~\bibnamefont {Meng}}, \bibinfo {author} {\bibfnamefont {X.-C.}\
  \bibnamefont {Ma}},\ and\ \bibinfo {author} {\bibfnamefont {Q.-K.}\
  \bibnamefont {Xue}},\ }\href {https://doi.org/10.1103/PhysRevB.101.085413}
  {\bibfield  {journal} {\bibinfo  {journal} {Phys. Rev. B}\ }\textbf {\bibinfo
  {volume} {101}},\ \bibinfo {pages} {85413} (\bibinfo {year}
  {2020})}\BibitemShut {NoStop}%
\bibitem [{\citenamefont {Ren}\ \emph {et~al.}(2020)\citenamefont {Ren},
  \citenamefont {Han}, \citenamefont {Wang}, \citenamefont {Fan}, \citenamefont
  {Song}, \citenamefont {Ma},\ and\ \citenamefont {Xue}}]{Ren2020}%
  \BibitemOpen
  \bibfield  {author} {\bibinfo {author} {\bibfnamefont {M.-Q.}\ \bibnamefont
  {Ren}}, \bibinfo {author} {\bibfnamefont {S.}~\bibnamefont {Han}}, \bibinfo
  {author} {\bibfnamefont {S.-Z.}\ \bibnamefont {Wang}}, \bibinfo {author}
  {\bibfnamefont {J.-Q.}\ \bibnamefont {Fan}}, \bibinfo {author} {\bibfnamefont
  {C.-L.}\ \bibnamefont {Song}}, \bibinfo {author} {\bibfnamefont {X.-C.}\
  \bibnamefont {Ma}},\ and\ \bibinfo {author} {\bibfnamefont {Q.-K.}\
  \bibnamefont {Xue}},\ }\href {https://doi.org/10.1103/PhysRevLett.124.187001}
  {\bibfield  {journal} {\bibinfo  {journal} {Phys. Rev. Lett.}\ }\textbf
  {\bibinfo {volume} {124}},\ \bibinfo {pages} {187001} (\bibinfo {year}
  {2020})}\BibitemShut {NoStop}%
\bibitem [{\citenamefont {Mitrano}\ \emph {et~al.}(2016)\citenamefont
  {Mitrano}, \citenamefont {Cantaluppi}, \citenamefont {Nicoletti},
  \citenamefont {Kaiser}, \citenamefont {Perucchi}, \citenamefont {Lupi},
  \citenamefont {Pietro}, \citenamefont {Pontiroli}, \citenamefont {Riccò},
  \citenamefont {Clark}, \citenamefont {Jaksch},\ and\ \citenamefont
  {Cavalleri}}]{Mitrano2016}%
  \BibitemOpen
  \bibfield  {author} {\bibinfo {author} {\bibfnamefont {M.}~\bibnamefont
  {Mitrano}}, \bibinfo {author} {\bibfnamefont {A.}~\bibnamefont {Cantaluppi}},
  \bibinfo {author} {\bibfnamefont {D.}~\bibnamefont {Nicoletti}}, \bibinfo
  {author} {\bibfnamefont {S.}~\bibnamefont {Kaiser}}, \bibinfo {author}
  {\bibfnamefont {A.}~\bibnamefont {Perucchi}}, \bibinfo {author}
  {\bibfnamefont {S.}~\bibnamefont {Lupi}}, \bibinfo {author} {\bibfnamefont
  {P.~D.}\ \bibnamefont {Pietro}}, \bibinfo {author} {\bibfnamefont
  {D.}~\bibnamefont {Pontiroli}}, \bibinfo {author} {\bibfnamefont
  {M.}~\bibnamefont {Riccò}}, \bibinfo {author} {\bibfnamefont {S.~R.}\
  \bibnamefont {Clark}}, \bibinfo {author} {\bibfnamefont {D.}~\bibnamefont
  {Jaksch}},\ and\ \bibinfo {author} {\bibfnamefont {A.}~\bibnamefont
  {Cavalleri}},\ }\href {https://doi.org/10.1038/nature16522} {\bibfield
  {journal} {\bibinfo  {journal} {Nature (London)}\ }\textbf {\bibinfo {volume}
  {530}},\ \bibinfo {pages} {461} (\bibinfo {year} {2016})}\BibitemShut
  {NoStop}%
\bibitem [{\citenamefont {Cantaluppi}\ \emph {et~al.}(2018)\citenamefont
  {Cantaluppi}, \citenamefont {Buzzi}, \citenamefont {Jotzu}, \citenamefont
  {Nicoletti}, \citenamefont {Mitrano}, \citenamefont {Pontiroli},
  \citenamefont {Riccò}, \citenamefont {Perucchi}, \citenamefont {Pietro},\
  and\ \citenamefont {Cavalleri}}]{Cantaluppi2018}%
  \BibitemOpen
  \bibfield  {author} {\bibinfo {author} {\bibfnamefont {A.}~\bibnamefont
  {Cantaluppi}}, \bibinfo {author} {\bibfnamefont {M.}~\bibnamefont {Buzzi}},
  \bibinfo {author} {\bibfnamefont {G.}~\bibnamefont {Jotzu}}, \bibinfo
  {author} {\bibfnamefont {D.}~\bibnamefont {Nicoletti}}, \bibinfo {author}
  {\bibfnamefont {M.}~\bibnamefont {Mitrano}}, \bibinfo {author} {\bibfnamefont
  {D.}~\bibnamefont {Pontiroli}}, \bibinfo {author} {\bibfnamefont
  {M.}~\bibnamefont {Riccò}}, \bibinfo {author} {\bibfnamefont
  {A.}~\bibnamefont {Perucchi}}, \bibinfo {author} {\bibfnamefont {P.~D.}\
  \bibnamefont {Pietro}},\ and\ \bibinfo {author} {\bibfnamefont
  {A.}~\bibnamefont {Cavalleri}},\ }\href
  {https://doi.org/10.1038/s41567-018-0134-8} {\bibfield  {journal} {\bibinfo
  {journal} {Nat. Phys.}\ }\textbf {\bibinfo {volume} {14}},\ \bibinfo {pages}
  {837} (\bibinfo {year} {2018})}\BibitemShut {NoStop}%
\bibitem [{\citenamefont {Capone}\ \emph {et~al.}(2000)\citenamefont {Capone},
  \citenamefont {Fabrizio}, \citenamefont {Giannozzi},\ and\ \citenamefont
  {Tosatti}}]{Capone2000}%
  \BibitemOpen
  \bibfield  {author} {\bibinfo {author} {\bibfnamefont {M.}~\bibnamefont
  {Capone}}, \bibinfo {author} {\bibfnamefont {M.}~\bibnamefont {Fabrizio}},
  \bibinfo {author} {\bibfnamefont {P.}~\bibnamefont {Giannozzi}},\ and\
  \bibinfo {author} {\bibfnamefont {E.}~\bibnamefont {Tosatti}},\ }\href
  {https://doi.org/10.1103/PhysRevB.62.7619} {\bibfield  {journal} {\bibinfo
  {journal} {Phys. Rev. B}\ }\textbf {\bibinfo {volume} {62}},\ \bibinfo
  {pages} {7619} (\bibinfo {year} {2000})}\BibitemShut {NoStop}%
\bibitem [{\citenamefont {Nomura}\ \emph {et~al.}(2015)\citenamefont {Nomura},
  \citenamefont {Sakai}, \citenamefont {Capone},\ and\ \citenamefont
  {Arita}}]{Nomura2015sci}%
  \BibitemOpen
  \bibfield  {author} {\bibinfo {author} {\bibfnamefont {Y.}~\bibnamefont
  {Nomura}}, \bibinfo {author} {\bibfnamefont {S.}~\bibnamefont {Sakai}},
  \bibinfo {author} {\bibfnamefont {M.}~\bibnamefont {Capone}},\ and\ \bibinfo
  {author} {\bibfnamefont {R.}~\bibnamefont {Arita}},\ }\href
  {https://doi.org/10.1126/sciadv.1500568} {\bibfield  {journal} {\bibinfo
  {journal} {Sci. Adv.}\ }\textbf {\bibinfo {volume} {1}},\ \bibinfo {pages}
  {e1500568} (\bibinfo {year} {2015})}\BibitemShut {NoStop}%
\bibitem [{\citenamefont {Koga}\ and\ \citenamefont {Werner}(2015)}]{Koga2015}%
  \BibitemOpen
  \bibfield  {author} {\bibinfo {author} {\bibfnamefont {A.}~\bibnamefont
  {Koga}}\ and\ \bibinfo {author} {\bibfnamefont {P.}~\bibnamefont {Werner}},\
  }\href {https://doi.org/10.1103/PhysRevB.91.085108} {\bibfield  {journal}
  {\bibinfo  {journal} {Phys. Rev. B}\ }\textbf {\bibinfo {volume} {91}},\
  \bibinfo {pages} {085108} (\bibinfo {year} {2015})}\BibitemShut {NoStop}%
\bibitem [{\citenamefont {Hoshino}\ and\ \citenamefont
  {Werner}(2016)}]{Hoshino2016}%
  \BibitemOpen
  \bibfield  {author} {\bibinfo {author} {\bibfnamefont {S.}~\bibnamefont
  {Hoshino}}\ and\ \bibinfo {author} {\bibfnamefont {P.}~\bibnamefont
  {Werner}},\ }\href {https://doi.org/10.1103/PhysRevB.93.155161} {\bibfield
  {journal} {\bibinfo  {journal} {Phys. Rev. B}\ }\textbf {\bibinfo {volume}
  {93}},\ \bibinfo {pages} {155161} (\bibinfo {year} {2016})}\BibitemShut
  {NoStop}%
\bibitem [{\citenamefont {Steiner}\ \emph {et~al.}(2016)\citenamefont
  {Steiner}, \citenamefont {Hoshino}, \citenamefont {Nomura},\ and\
  \citenamefont {Werner}}]{Steiner2016}%
  \BibitemOpen
  \bibfield  {author} {\bibinfo {author} {\bibfnamefont {K.}~\bibnamefont
  {Steiner}}, \bibinfo {author} {\bibfnamefont {S.}~\bibnamefont {Hoshino}},
  \bibinfo {author} {\bibfnamefont {Y.}~\bibnamefont {Nomura}},\ and\ \bibinfo
  {author} {\bibfnamefont {P.}~\bibnamefont {Werner}},\ }\href
  {https://doi.org/10.1103/PhysRevB.94.075107} {\bibfield  {journal} {\bibinfo
  {journal} {Phys. Rev. B}\ }\textbf {\bibinfo {volume} {94}},\ \bibinfo
  {pages} {075107} (\bibinfo {year} {2016})}\BibitemShut {NoStop}%
\bibitem [{\citenamefont {Hoshino}\ and\ \citenamefont
  {Werner}(2017)}]{Hoshino2017}%
  \BibitemOpen
  \bibfield  {author} {\bibinfo {author} {\bibfnamefont {S.}~\bibnamefont
  {Hoshino}}\ and\ \bibinfo {author} {\bibfnamefont {P.}~\bibnamefont
  {Werner}},\ }\href {https://doi.org/10.1103/PhysRevLett.118.177002}
  {\bibfield  {journal} {\bibinfo  {journal} {Phys. Rev. Lett.}\ }\textbf
  {\bibinfo {volume} {118}},\ \bibinfo {pages} {177002} (\bibinfo {year}
  {2017})}\BibitemShut {NoStop}%
\bibitem [{\citenamefont {Ishigaki}\ \emph {et~al.}(2018)\citenamefont
  {Ishigaki}, \citenamefont {Nasu}, \citenamefont {Koga}, \citenamefont
  {Hoshino},\ and\ \citenamefont {Werner}}]{Ishigaki2018}%
  \BibitemOpen
  \bibfield  {author} {\bibinfo {author} {\bibfnamefont {K.}~\bibnamefont
  {Ishigaki}}, \bibinfo {author} {\bibfnamefont {J.}~\bibnamefont {Nasu}},
  \bibinfo {author} {\bibfnamefont {A.}~\bibnamefont {Koga}}, \bibinfo {author}
  {\bibfnamefont {S.}~\bibnamefont {Hoshino}},\ and\ \bibinfo {author}
  {\bibfnamefont {P.}~\bibnamefont {Werner}},\ }\href
  {https://doi.org/10.1103/PhysRevB.98.235120} {\bibfield  {journal} {\bibinfo
  {journal} {Phys. Rev. B}\ }\textbf {\bibinfo {volume} {98}},\ \bibinfo
  {pages} {235120} (\bibinfo {year} {2018})}\BibitemShut {NoStop}%
\bibitem [{\citenamefont {Ishigaki}\ \emph {et~al.}(2019)\citenamefont
  {Ishigaki}, \citenamefont {Nasu}, \citenamefont {Koga}, \citenamefont
  {Hoshino},\ and\ \citenamefont {Werner}}]{Ishigaki2019}%
  \BibitemOpen
  \bibfield  {author} {\bibinfo {author} {\bibfnamefont {K.}~\bibnamefont
  {Ishigaki}}, \bibinfo {author} {\bibfnamefont {J.}~\bibnamefont {Nasu}},
  \bibinfo {author} {\bibfnamefont {A.}~\bibnamefont {Koga}}, \bibinfo {author}
  {\bibfnamefont {S.}~\bibnamefont {Hoshino}},\ and\ \bibinfo {author}
  {\bibfnamefont {P.}~\bibnamefont {Werner}},\ }\href
  {https://doi.org/10.1103/PhysRevB.99.085131} {\bibfield  {journal} {\bibinfo
  {journal} {Phys. Rev. B}\ }\textbf {\bibinfo {volume} {99}},\ \bibinfo
  {pages} {085131} (\bibinfo {year} {2019})}\BibitemShut {NoStop}%
\bibitem [{\citenamefont {Yue}\ \emph {et~al.}(2020)\citenamefont {Yue},
  \citenamefont {Hoshino},\ and\ \citenamefont {Werner}}]{Yue2020nov}%
  \BibitemOpen
  \bibfield  {author} {\bibinfo {author} {\bibfnamefont {C.}~\bibnamefont
  {Yue}}, \bibinfo {author} {\bibfnamefont {S.}~\bibnamefont {Hoshino}},\ and\
  \bibinfo {author} {\bibfnamefont {P.}~\bibnamefont {Werner}},\ }\href
  {https://doi.org/10.1103/PhysRevB.102.195103} {\bibfield  {journal} {\bibinfo
   {journal} {Phys. Rev. B}\ }\textbf {\bibinfo {volume} {102}},\ \bibinfo
  {pages} {195103} (\bibinfo {year} {2020})}\BibitemShut {NoStop}%
\bibitem [{\citenamefont {Hoshino}\ \emph {et~al.}(2019)\citenamefont
  {Hoshino}, \citenamefont {Werner},\ and\ \citenamefont
  {Arita}}]{Hoshino2019}%
  \BibitemOpen
  \bibfield  {author} {\bibinfo {author} {\bibfnamefont {S.}~\bibnamefont
  {Hoshino}}, \bibinfo {author} {\bibfnamefont {P.}~\bibnamefont {Werner}},\
  and\ \bibinfo {author} {\bibfnamefont {R.}~\bibnamefont {Arita}},\ }\href
  {https://doi.org/10.1103/PhysRevB.99.235133} {\bibfield  {journal} {\bibinfo
  {journal} {Phys. Rev. B}\ }\textbf {\bibinfo {volume} {99}},\ \bibinfo
  {pages} {235133} (\bibinfo {year} {2019})}\BibitemShut {NoStop}%
\bibitem [{\citenamefont {Misawa}\ and\ \citenamefont
  {Imada}(2017)}]{Misawa2017}%
  \BibitemOpen
  \bibfield  {author} {\bibinfo {author} {\bibfnamefont {T.}~\bibnamefont
  {Misawa}}\ and\ \bibinfo {author} {\bibfnamefont {M.}~\bibnamefont {Imada}},\
  }\href {https://arxiv.org/abs/1711.10205} {\bibfield  {journal} {\bibinfo
  {journal} {arXiv:1711.10205}\ } (\bibinfo {year} {2017})}\BibitemShut
  {NoStop}%
\bibitem [{\citenamefont {Kugel}\ and\ \citenamefont
  {Khomskii}(1972)}]{Kugel72}%
  \BibitemOpen
  \bibfield  {author} {\bibinfo {author} {\bibfnamefont {K.~I.}\ \bibnamefont
  {Kugel}}\ and\ \bibinfo {author} {\bibfnamefont {D.~I.}\ \bibnamefont
  {Khomskii}},\ }\href@noop {} {\bibfield  {journal} {\bibinfo  {journal}
  {ZhETF Pis. Red.}\ }\textbf {\bibinfo {volume} {15}},\ \bibinfo {pages} {629}
  (\bibinfo {year} {1972})},\ \bibinfo {note} {[Sov. Phys. JETP Lett.
  \textbf{15}, 446 (1972)]}\BibitemShut {NoStop}%
\bibitem [{\citenamefont {Kugel}\ and\ \citenamefont
  {Khomskii}(1973)}]{Kugel73}%
  \BibitemOpen
  \bibfield  {author} {\bibinfo {author} {\bibfnamefont {K.~I.}\ \bibnamefont
  {Kugel}}\ and\ \bibinfo {author} {\bibfnamefont {D.~I.}\ \bibnamefont
  {Khomskii}},\ }\href@noop {} {\bibfield  {journal} {\bibinfo  {journal} {Zh.
  Eksp. Teor. Fiz.}\ }\textbf {\bibinfo {volume} {64}},\ \bibinfo {pages}
  {1429} (\bibinfo {year} {1973})},\ \bibinfo {note} {[Sov. Phys. JETP
  \textbf{37}, 725 (1973)]}\BibitemShut {NoStop}%
\bibitem [{\citenamefont {Inagaki}(1975)}]{Inagaki1975}%
  \BibitemOpen
  \bibfield  {author} {\bibinfo {author} {\bibfnamefont {S.}~\bibnamefont
  {Inagaki}},\ }\href {https://doi.org/10.1143/JPSJ.39.596} {\bibfield
  {journal} {\bibinfo  {journal} {J. Phys. Soc. Jpn.}\ }\textbf {\bibinfo
  {volume} {39}},\ \bibinfo {pages} {596} (\bibinfo {year} {1975})}\BibitemShut
  {NoStop}%
\bibitem [{\citenamefont {Ishihara}\ \emph {et~al.}(1997)\citenamefont
  {Ishihara}, \citenamefont {Inoue},\ and\ \citenamefont
  {Maekawa}}]{Ishihara1997}%
  \BibitemOpen
  \bibfield  {author} {\bibinfo {author} {\bibfnamefont {S.}~\bibnamefont
  {Ishihara}}, \bibinfo {author} {\bibfnamefont {J.}~\bibnamefont {Inoue}},\
  and\ \bibinfo {author} {\bibfnamefont {S.}~\bibnamefont {Maekawa}},\ }\href
  {https://doi.org/10.1103/PhysRevB.55.8280} {\bibfield  {journal} {\bibinfo
  {journal} {Phys. Rev. B}\ }\textbf {\bibinfo {volume} {55}},\ \bibinfo
  {pages} {8280} (\bibinfo {year} {1997})}\BibitemShut {NoStop}%
\bibitem [{\citenamefont {Feiner}\ and\ \citenamefont
  {Oleś}(1999)}]{Feiner1999}%
  \BibitemOpen
  \bibfield  {author} {\bibinfo {author} {\bibfnamefont {L.~F.}\ \bibnamefont
  {Feiner}}\ and\ \bibinfo {author} {\bibfnamefont {A.~M.}\ \bibnamefont
  {Oleś}},\ }\href {https://doi.org/10.1103/PhysRevB.59.3295} {\bibfield
  {journal} {\bibinfo  {journal} {Phys. Rev. B}\ }\textbf {\bibinfo {volume}
  {59}},\ \bibinfo {pages} {3295} (\bibinfo {year} {1999})}\BibitemShut
  {NoStop}%
\bibitem [{\citenamefont {Normand}\ and\ \citenamefont
  {Oleś}(2008)}]{Normand2008}%
  \BibitemOpen
  \bibfield  {author} {\bibinfo {author} {\bibfnamefont {B.}~\bibnamefont
  {Normand}}\ and\ \bibinfo {author} {\bibfnamefont {A.~M.}\ \bibnamefont
  {Oleś}},\ }\href {https://doi.org/10.1103/PhysRevB.78.094427} {\bibfield
  {journal} {\bibinfo  {journal} {Phys. Rev. B}\ }\textbf {\bibinfo {volume}
  {78}},\ \bibinfo {pages} {094427} (\bibinfo {year} {2008})}\BibitemShut
  {NoStop}%
\bibitem [{\citenamefont {Tosatti}\ \emph {et~al.}(1996)\citenamefont
  {Tosatti}, \citenamefont {Manini},\ and\ \citenamefont
  {Gunnarsson}}]{Tosatti1996}%
  \BibitemOpen
  \bibfield  {author} {\bibinfo {author} {\bibfnamefont {E.}~\bibnamefont
  {Tosatti}}, \bibinfo {author} {\bibfnamefont {N.}~\bibnamefont {Manini}},\
  and\ \bibinfo {author} {\bibfnamefont {O.}~\bibnamefont {Gunnarsson}},\
  }\href {https://doi.org/10.1103/PhysRevB.54.17184} {\bibfield  {journal}
  {\bibinfo  {journal} {Phys. Rev. B}\ }\textbf {\bibinfo {volume} {54}},\
  \bibinfo {pages} {17184} (\bibinfo {year} {1996})}\BibitemShut {NoStop}%
\bibitem [{\citenamefont {Nomura}\ and\ \citenamefont
  {Arita}(2015)}]{Nomura2015prb}%
  \BibitemOpen
  \bibfield  {author} {\bibinfo {author} {\bibfnamefont {Y.}~\bibnamefont
  {Nomura}}\ and\ \bibinfo {author} {\bibfnamefont {R.}~\bibnamefont {Arita}},\
  }\href {https://doi.org/10.1103/PhysRevB.92.245108} {\bibfield  {journal}
  {\bibinfo  {journal} {Phys. Rev. B}\ }\textbf {\bibinfo {volume} {92}},\
  \bibinfo {pages} {245108} (\bibinfo {year} {2015})}\BibitemShut {NoStop}%
\bibitem [{\citenamefont {Nomura}\ \emph {et~al.}(2012)\citenamefont {Nomura},
  \citenamefont {Nakamura},\ and\ \citenamefont {Arita}}]{Nomura2012}%
  \BibitemOpen
  \bibfield  {author} {\bibinfo {author} {\bibfnamefont {Y.}~\bibnamefont
  {Nomura}}, \bibinfo {author} {\bibfnamefont {K.}~\bibnamefont {Nakamura}},\
  and\ \bibinfo {author} {\bibfnamefont {R.}~\bibnamefont {Arita}},\ }\href
  {https://doi.org/10.1103/PhysRevB.85.155452} {\bibfield  {journal} {\bibinfo
  {journal} {Phys. Rev. B}\ }\textbf {\bibinfo {volume} {85}},\ \bibinfo
  {pages} {155452} (\bibinfo {year} {2012})}\BibitemShut {NoStop}%
\bibitem [{\citenamefont {Kasahara}\ \emph {et~al.}(2014)\citenamefont
  {Kasahara}, \citenamefont {Takeuchi}, \citenamefont {Itou}, \citenamefont
  {Zadik}, \citenamefont {Takabayashi}, \citenamefont {Ganin}, \citenamefont
  {Arčon}, \citenamefont {Rosseinsky}, \citenamefont {Prassides},\ and\
  \citenamefont {Iwasa}}]{Kasahara2014}%
  \BibitemOpen
  \bibfield  {author} {\bibinfo {author} {\bibfnamefont {Y.}~\bibnamefont
  {Kasahara}}, \bibinfo {author} {\bibfnamefont {Y.}~\bibnamefont {Takeuchi}},
  \bibinfo {author} {\bibfnamefont {T.}~\bibnamefont {Itou}}, \bibinfo {author}
  {\bibfnamefont {R.~H.}\ \bibnamefont {Zadik}}, \bibinfo {author}
  {\bibfnamefont {Y.}~\bibnamefont {Takabayashi}}, \bibinfo {author}
  {\bibfnamefont {A.~Y.}\ \bibnamefont {Ganin}}, \bibinfo {author}
  {\bibfnamefont {D.}~\bibnamefont {Arčon}}, \bibinfo {author} {\bibfnamefont
  {M.~J.}\ \bibnamefont {Rosseinsky}}, \bibinfo {author} {\bibfnamefont
  {K.}~\bibnamefont {Prassides}},\ and\ \bibinfo {author} {\bibfnamefont
  {Y.}~\bibnamefont {Iwasa}},\ }\href
  {https://doi.org/10.1103/PhysRevB.90.014413} {\bibfield  {journal} {\bibinfo
  {journal} {Phys. Rev. B}\ }\textbf {\bibinfo {volume} {90}},\ \bibinfo
  {pages} {014413} (\bibinfo {year} {2014})}\BibitemShut {NoStop}%
\end{thebibliography}%

\end{document}